\documentclass{elsart}
\usepackage[centertags]{amsmath}
\usepackage{bbm}

\numberwithin{equation}{section}

\DeclareMathOperator{\re}{Re}
\DeclareMathOperator{\im}{Im}
\DeclareMathOperator{\tr}{tr}
\DeclareMathOperator{\diag}{diag}

\newcommand{\cba}{\bar c}
\newcommand{\dba}{\bar d}
\newcommand{\1}{\mathbbm{1}}
\newcommand{\Zd}{Z^\dagger}
\newcommand{\nn}{\notag}

\begin{document}

\begin{frontmatter}

\title{Bosonic color-flavor transformation for the special unitary group}

\author[Yale,Regensburg]{Yi Wei} and
\author[Regensburg]{Tilo Wettig}
\address[Yale]{Department of Physics, Yale University,
  New Haven, CT 06520-8120, USA}
\address[Regensburg]{Institute for Theoretical Physics, University of
  Regensburg, 93040 Regensburg, Germany}
\date{26 November 2004}

\begin{abstract}
  We extend Zirnbauer's color-flavor transformation in the bosonic
  sector to the color group SU$(N_c)$.  Because the flavor group
  U$(N_b, N_b)$ is non-compact, the algebraic method by which the
  original color-flavor transformation was derived leads to a useful
  result only for $2N_b\leq N_c$.  Using the character expansion
  method, we obtain a different form of the transformation in the
  extended range $N_b\leq N_c$.  This result can also be used for the
  color group U$(N_c)$.  The integrals to which the transformation can
  be applied are of relevance for the recently proposed boson-induced
  lattice gauge theory.
\end{abstract}

\end{frontmatter}

\section{Introduction}

In 1996, Zirnbauer~\cite{Zirnbauer:1996} invented a generalized
Hubbard-Stratonovitch transformation which trades an integration over
a ``color'' gauge group for an integration over a certain
supersymmetric coset space, or ``flavor'' space.  Although the
transformation was originally derived to study disordered systems in
condensed matter physics, the terminology comes from lattice gauge
theory because the integral over the gauge group to which the
color-flavor transformation is applied is precisely of the form of a
one-link integral in lattice gauge theory at infinite coupling.

The fields that appear in the transformation carry two types of
indices that will be refered to as color and flavor indices.  The
number of colors is denoted by $N_c$, and the numbers of bosonic and
fermionic flavors are denoted by $N_b$ and $N_f$, respectively.
Zirnbauer derived versions of the color-flavor transformation for the
three color groups U($N_c$), O($N_c$), and
Sp($2N_c$)~\cite{Zirnbauer:1996}.  In his original work, the flavor
space contained an equal number of bosonic and fermionic degrees of
freedom, but it is possible to relax this
constraint~\cite{Zirnbauer:1997qm}.  Convergence requirements then
place an upper bound on the difference between the number of bosons
and fermions.  For the case of U$(N_c)$, this bound is given by
$2(N_b-N_f)\leq N_c$.

The color-flavor transformation has been used in a number of physical
applications, e.g., in the derivation of a field theory for the random
flux model by Altland and Simons~\cite{Altl99} and in the construction
of chiral Lagrangians for lattice gauge theories by Nagao and
Nishigaki~\cite{Nagao:2000ke}.  In the latter paper, the calculations
were done for the above-mentioned color groups.  However, in quantum
chromodynamics (QCD) the color group is SU(3).  To be able to apply
the color-flavor transformation to this very important physical case,
a variant of the transformation for the special unitary group needed
to be derived.  Following earlier work by Budczies and
Shnir~\cite{Budczies:2000qs,Budczies:2000a}, this was done in
Refs.~\cite{fermi-cft,BNSZ} in the fermionic sector, i.e. for $N_b=0$.
The result was then applied to lattice QCD in
Refs.~\cite{BNSZ,cft-app}.

As mentioned above, the color-flavor transformation can only be
applied to the one-link integral of lattice gauge theory if the gauge
coupling is infinite.  Fortunately, it is possible to go beyond the
infinite-coupling limit.  A gauge action (Yang-Mills action in the
continuum or Wilson action on the lattice) can be induced by coupling
auxiliary fields to the gauge field (still at infinite coupling) and
integrating out these extra fields.  This idea, which is known as
``induced QCD'', has been considered in various forms in the
literature \cite{Bander,Hamber,KM,HH,MZ,NS}.  For example, the job can
be done by a number $N_h$ of heavy auxiliary fermions with common mass
$m_h$ in the combined limit $N_h\to\infty$, $m_h\to\infty$ such that
the ratio $N_h/m_h^4$ is constant, and this constant can then be
related to the strength of the induced gauge coupling.

So far, the color-flavor transformation for SU$(N_c)$ has only been
derived in the fermionic sector, for three reasons: (a) physical
quarks are fermions, (b) a gauge action can be induced by auxiliary
fermions alone, and (c) the calculation is somewhat easier for
fermions than for bosons.  There seemed to be no special need for a
bosonic variant of the transformation until Budczies and Zirnbauer
suggested a new method to induce the gauge action using auxiliary
bosons~\cite{induce}.  Their approach has the advantage of requiring
only a small number (essentially equal to $N_c$) of auxiliary fields.
However, to be able to use their method in lattice QCD, one requires
an SU($N_c$)-variant of the color-flavor transformation that
accommodates both fermions (the physical quarks) and bosons (to induce
the gauge action).

As a first step towards this goal, we found it useful to consider the
purely bosonic case with $N_f=0$ for which we will present results in
this paper.  Our results can be applied to study a boson-induced
gauge theory without physical quarks.  The concentration on the
bosonic sector allows us to separate the complications due to the fact
that the fields are bosonic from those due to the supersymmetric
framework.  This is the main motivation for the present paper.  The
supersymmetric case will be addressed in a separate publication.

The convergence requirement mentioned above is irrelevant for the
purely fermionic case in which $N_b=0$ since the inequality is always
satisfied.  However, it becomes relevant for the purely bosonic case.
We obtain a ``standard'' form of the color-flavor transformation if
the condition $2N_b\leq N_c$ is satisfied.  In the extended range
$N_b\leq N_c$ we derive a different form of the transformation.
Interestingly, for $N_b<N_c$ our results for SU($N_c$) are identical
to those for gauge group U($N_c$), and we therefore obtain new results
for U($N_c$) in the range $N_b<N_c<2N_b$.  For $N_b>N_c$ we have not
been able to simplify our formal result to be useful in applications.

This paper is organized as follows. We first state our results in
Sec.~\ref{sec:results}.  In Sec.~\ref{sec:alg}, we use the algebraic
method of Refs.~\cite{Zirnbauer:1996,fermi-cft} to derive our general
result for the bosonic color-flavor transformation.  In
Sec.~\ref{sec:char}, we use a different approach, the character
expansion method, to derive an alternative form of the bosonic
color-flavor transformation.  We close with a brief discussion of
possible applications and open problems.  The appendix contains
derivations of several intermediate results as well as a number of
examples for the results from both approaches.

\section{Statement of results}
\label{sec:results}

Let $\psi_a^i$, $\bar \psi_a^i$, $\varphi_a^i$, and $\bar\varphi_a^i$
be complex bosonic variables that carry a color index (superscript $i$
running from 1 to $N_c$) and a flavor index (subscript $a$ running
from 1 to $N_b$).  The bar denotes complex conjugation.  Summation
over repeated indices is implied here and throughout the paper unless
indicated otherwise.  Using the algebraic method of
Refs.~\cite{Zirnbauer:1996,fermi-cft}, we obtain
\begin{align}
  \label{eq:cft-alg}
  \int_{{\rm SU}(N_c)} & dU \exp
  \left(\bar{\psi}_{a}^{i}U^{ij}\psi_{a}^{j}+
    \bar{\varphi}_{a}^{i}U^{\dagger ij}\varphi_{a}^{j} \right)\\
  &=\int_{|Z\Zd|\leq 1}\frac{dZd\Zd}{{\det}^{2N_b-N_c}(1-ZZ^{\dagger})}
  \exp\left(\bar{\psi}_a^i Z_{ab}\varphi_b^i
    +\bar{\varphi}_a^i\Zd_{ab}\psi_b^i \right) \sum_{Q=0}^{\infty}
  \chi_Q \nn\:,
\end{align}
where 
\begin{equation}
  \label{eq:chiQ}
  \chi_0=\mathcal C_0\:, \qquad\chi_{Q>0}= \mathcal{C}_Q \left(
    {\det}^Q\mathcal{M}+{\det}^Q\mathcal{N}\right) \:,
\end{equation}
the $N_c \times N_c$ matrices $\mathcal M$ and $\mathcal N$ are
defined by ${\mathcal M}^{ij}={\bar \psi}^i_a(1-Z\Zd)_{ab}\psi^j_b$
and ${\mathcal N}^{ij}={\bar \varphi}^i_a(1-\Zd Z)_{ab}\varphi^j_b$,
and the coefficients $\mathcal C_Q$ are computed in
App.~\ref{app:rad}.

The integration on the left-hand side (LHS) of Eq.~\eqref{eq:cft-alg}
is over SU($N_c$) matrices $U$ distributed according to the Haar
measure $dU$, normalized such that the group volume is unity.  The
integration on the right-hand side (RHS) of that equation is over
complex $N_b\times N_b$ matrices $Z$, with the restriction that all
eigenvalues of $Z\Zd$ are less than or equal to one.  These matrices
parameterize the non-compact coset space $\mathrm U(N_b,N_b)/[\mathrm
U(N_b)\times \mathrm U(N_b)]$.  The corresponding invariant
integration measure is given by \cite{hua}
\begin{equation}
  \label{eq:dz}
  d\mu(Z,Z^{\dag})=\frac{dZdZ^{\dag}}{{\det}(1-ZZ^{\dag})^{
  2N_b}}\quad\text{with}\quad
  dZdZ^{\dag}=\prod^{N_b}_{a,b=1}d\re Z_{ab}\ d\im Z_{ab}\:.
\end{equation}

Note that the $N_c\times N_c$ matrices $\mathcal M$ and $\mathcal N$
can be thought of as products of three matrices of dimension
$N_c\times N_b$, $N_b\times N_b$, and $N_b\times N_c$, respectively.
The resulting matrix is of full rank only if $N_b\geq N_c$.  For
$N_b<N_c$, we therefore have $\det\mathcal M=\det\mathcal N=0$
\cite{lc}, and the transformation simplifies to
\begin{align}
  \label{eq:cft-simple}
  \int_{{\rm SU}(N_c)} & dU
    \exp\left(\bar{\psi}_{a}^{i}U^{ij}\psi_{a}^{j}+ 
    \bar{\varphi}_{a}^{i}U^{\dagger ij}\varphi_{a}^{j} \right)\nn \\
  &=\mathcal C_0 \int_{|Z\Zd|\leq 1}
  \frac{dZd\Zd}{{\det}^{2N_b-N_c}(1-ZZ^{\dagger})}
  \exp\left(\bar{\psi}_a^i Z_{ab}\varphi_b^i
    +\bar{\varphi}_a^i\Zd_{ab}\psi_b^i \right) 
\end{align}
with a constant $\mathcal C_0$ given in Eq.~\eqref{eq:C0}.  This
agrees with the result for the color group U($N_c$) in
Ref.~\cite{Zirnbauer:1997qm}.

Equation~\eqref{eq:cft-alg} looks similar to the result for the
fermionic sector presented in Ref.~\cite{fermi-cft}.  There are two
major differences, however.  First, due to the nilpotency of Grassmann
variables the sum over $Q$ in Eq.~\eqref{eq:cft-alg} only went up to
$N_f$ in the fermionic case, whereas it extends to infinity now.
Second, the invariant measure of the coset space $\mathrm {U}(N_b,
N_b)/[\mathrm {U}(N_b)\times \mathrm {U}(N_b)]$ in Eq.~\eqref{eq:dz}
diverges at the boundary, giving rise to convergence issues which we
discuss now.

For $2N_b\leq N_c$, the divergence of the measure is canceled by the
factor of ${\det}^{N_c}(1-ZZ^{\dagger})$ in the integrand of
Eq.~\eqref{eq:cft-alg}.  In this case the result \eqref{eq:cft-simple}
applies and is free from divergences.  For $N_b<N_c<2N_b$, the
integral over $Z$ in Eq.~\eqref{eq:cft-simple} diverges.  For
$N_b=N_c$, the integral over $Z$ in Eq.~\eqref{eq:cft-alg} diverges for
$Q<N_b$, whereas for $N_b>N_c$, it diverges for all $Q$.  Of course,
the final result for the RHS of Eqs.~\eqref{eq:cft-alg} and
\eqref{eq:cft-simple} must be finite, so whatever divergence arises
from the integration over $Z$ will be canceled by a similar divergence
in the integral for the corresponding (inverse) constant $\mathcal
C_Q^{-1}$, see Eq.~\eqref{eq:radial-int}.  A finite ratio could in
principle be obtained by a limiting procedure, but it is not clear to
us whether this would lead to a simple final result.

Instead, we have used the character expansion
method~\cite{balant1,balant2,ce} to derive a different form of the
color-flavor transformation and obtain for $N_b<N_c$
\begin{align} 
  \label{eq:nf<nc}
  &\int_{{\rm SU}(N_c)} dU
  \exp\left(\bar{\psi}_{a}^{i}U^{ij}\psi_{a}^{j}+
    \bar{\varphi}_{a}^{i}U^{\dagger ij}\varphi_{a}^{j} \right) \\
  &=\prod_{n=0}^{N_c-N_b-1}\frac{(N_b+n)!}{n!}
  \int_{{\rm U}(N_b)} dV \ {\det}^{N_b-N_c}(VB)\exp 
  \left(\bar{\psi}^i_aV_{ab}\varphi^i_b
    +\bar{\varphi}^i_aV^\dagger_{ab}\psi^i_b\right)\:,\nn
\end{align} 
where the $N_b\times N_b$ matrix $B$ is defined by
$B_{ab}=\varphi_a^i{\bar \psi}^i_b$.  Note that the integration on the
RHS is over the unitary group with the normalized Haar
measure $dV$.  Equation~\eqref{eq:nf<nc} is also valid if the
integration on the LHS is over the color group U($N_c$) and, to the
best of our knowledge, represents a new result for this case.

The corresponding result for $N_b=N_c$ reads
\begin{align}
  \label{eq:nf=nc}
  \int_{{\rm SU}(N_c)} & dU 
  \exp\left(\bar{\psi}_{a}^{i}U^{ij}\psi_{a}^{j}+
    \bar{\varphi}_{a}^{i}U^{\dagger ij}\varphi_{a}^{j} \right) \nn \\ 
  &=\sum_{Q=0}^{\infty}\tilde\chi_Q\int_{{\rm U}(N_b)}
  dV \ {\det}^{-Q}(VB)\exp \left(\bar{\psi}^i_aV_{ab}\varphi^i_b
    +\bar{\varphi}^i_aV^\dagger_{ab}\psi^i_b\right)
\end{align}
with
\begin{equation}
  \label{eq:chi}
  \tilde\chi_0=1\:, \qquad \tilde\chi_{Q>0}={\det}^QM+{\det}^QN \:,
\end{equation}
the matrix $B$ as defined above, and $N_c\times N_c$ matrices $M$ and
$N$ defined by $M^{ij}=\psi^i_a\bar {\psi}_{a}^{j}$ and
$N^{ij}=\varphi_{a}^{i}\bar \varphi_{a}^{j}$.  Note that we are not
allowed to change the order of summation and integration in
Eq.~\eqref{eq:nf=nc}, see Sec.~\ref{sec:nf=nc}.  If the integration on
the LHS of Eq.~\eqref{eq:nf=nc} is over U($N_c$), only the $Q=0$ term
contributes on the RHS, see Eq.~\eqref{eq:Uint}.  The integral over
U($N_b$) in Eqs.~\eqref{eq:nf<nc} and \eqref{eq:nf=nc} can be done
analytically \cite{BRT,JSV,AW}, resulting in a determinant involving
modified Bessel functions, but we do not display this result here
because Eqs.~\eqref{eq:nf<nc} and \eqref{eq:nf=nc} are to be viewed as
transformations. 

As mentioned in the introduction, for $N_b>N_c$ we have not been able
to obtain a simple form of the color-flavor transformation in which
the divergences have been eliminated.

\section{Bosonic color-flavor transformation: algebraic method}
\label{sec:alg}

The basic idea of the algebraic approach to the color-flavor
transformation is to construct two projection operators onto the
subspace of Fock space (to be defined below) which is invariant under
the action of the color group $\mathrm {SU}(N_c)$.  One such projector
is implemented by integrating over the color group.  The other one is
obtained by integrating over a certain coset of the flavor group
U($N_b,N_b$).  Identification of the two projection operations then
leads to Eq.~\eqref{eq:cft-alg}.  In this section, we shall use this
algebraic approach to derive the bosonic color-flavor transformation.
We closely follow Refs.~\cite{Zirnbauer:1996,fermi-cft,BNSZ} whenever
possible.

\subsection{Fock space, Lie algebras, and Lie groups}

We introduce two sets of bosonic creation and annihilation operators
$\{ {\bar {c}}^{i}_{a}, {{c}}^{i}_{a} \}$ and $\{ {\bar d}^{i}_{a},
d^{i}_{a} \} $, where $i=1, \ldots, N_c$ and $a=1, \ldots, N_b$.  As
mentioned above, we shall refer to the upper index as ``color'' and to
the lower index as ``flavor''.  The Fock vacuum $|0\rangle$ is defined
by the requirement that $c^i_a|0\rangle=d^i_a|0\rangle=0$ for all
combinations of $i$ and $a$, and the Fock space is generated by acting
on $|0\rangle$ with the $\cba^i_a$ and $\dba^i_a$.  In the following,
the two sets of particles created by the $\cba^i_a$ and $\dba^i_a$
will be refered to as particles and antiparticles, respectively.

For simplicity of notation, we also introduce the unified operators $
\{ {\bar c}^{i}_{A}, c^{i}_{A} \}$ defined by
\begin{equation}
  \begin{split}
    c_A^i&=\left \{
      \begin{array}{ll}
        c_A^i & \text{for } 1\leq A\leq N_b \:, \\
        {\bar d}_{A-N_b}^i & \text{for } N_b < A\leq 2N_b \:,
      \end{array} \right. \\
    {\bar c}_A^i&=\left \{
      \begin{array}{ll}
        {\bar c}_A^i & \text{for } 1\leq A\leq N_b \:, \\
        {-d}_{A-N_b}^i & \text{for } N_b < A\leq 2N_b \:.
      \end{array} \right.
  \end{split}
\end{equation}
They satisfy the usual commutation relations for bosonic operators,
\begin{equation}
\label{eq:comm}
  [c^i_A, {\bar c}_B^j]=\delta^{ij} \delta_{AB}\:.
\end{equation}
Next we define operators $E^{ij}_{AB}={\bar c}^i_Ac^j_B$.  Simple
algebra reveals that they satisfy the commutation relations
\begin{equation}
  [E_{AB}^{ij},E_{CD}^{k\ell}]=\delta_{BC}\delta^{kj}E_{AD}^{i\ell}
  -\delta_{AD}\delta^{i\ell}E_{CB}^{kj}\:,
\end{equation} 
and hence they are generators of the Lie algebra $\mathrm {gl}
(2N_cN_b)$. 

The Lie algebra $\mathrm {gl} (2N_cN_b)$ has two commuting
subalgebras that are important for our proof, namely ${\mathrm {gl}}
(2N_b)$, which is generated by the color-singlet operators
\begin{align}
  \Bigl\{\mathcal{G}_{AB}\equiv \sum_{i=1}^{N_c}E_{AB}^{ii}\Bigr\}\:,
\end{align}
and ${\mathrm {sl}} (N_c)$, which is generated by the flavor-singlet
operators
\begin{equation}
  \label{eq:flsing}
  \begin{split}
    \Bigl\{\mathcal{E}^{ij}&\equiv
      \sum_{A=1}^{2N_b}E_{AA}^{ij};\:i\neq j\Bigr\} \quad\text{and} \\ 
    \Bigl\{\mathcal{H}^i&\equiv\sum_{A=1}^{2N_b}E_{AA}^{ii}-\frac
      {1}{N_c}\sum_{j=1}^{N_c}\sum_{A=1}^{2N_b}E_{AA}^{jj};\:
      i=1,\ldots,N_c\Bigr\}\:.
  \end{split}
\end{equation}
Note that only $N_c-1$ of the generators $\mathcal H^i$ are linearly
independent.

The action of the group GL$(2N_cN_b)$ and its subgroups on the Fock
space is defined by exponential mapping, i.e. for all $g \in
\text{GL}(2N_cN_b)$ we define a map $g\mapsto T_g$ from group elements
to operators by~\cite{Zirnbauer:1996,BNSZ}
\begin{equation}
  \label{eq:tg}  
  T_g=\exp \left[{\cba}^i_A(\ln g)_{AB}^{ij}c^j_B\right]\:.
\end{equation}
Following Zirnbauer~\cite{Zirnbauer:1996}, one can show that the map
$g\mapsto T_g$ is well-defined and a homomorphism of GL$(2N_cN_b)$,
\begin{equation}
  \label{eq:homo}
  T_g T_h=T_{gh}\:.
\end{equation}
Therefore it furnishes a (reducible) representation of GL$(2N_cN_b)$.

In the following, we will consider the action of the subgroups
SU($N_c$) (the color group) and U($N_b,N_b$) (the flavor group) of
GL$(2N_cN_b)$ on the Fock space.  The corresponding subalgebras
${\mathrm {sl}} (N_c)$ and ${\mathrm {gl}} (2N_b)$ have been given
above.  What are the reasons to single out these two subgroups?  For
the color group the reason is simple: The integration on the LHS of
Eq.~\eqref{eq:cft-alg} is over SU($N_c$).  For the flavor group, the
choice of the non-compact subgroup U($N_b,N_b$) is not immediately
obvious at this point but will become clear as we proceed.  We shall
see below that the color-neutral sector of Fock space is non-compact
in the bosonic case, as opposed to the fermionic case in which it was
compact.  Attempts to work with the compact subgroup U($2N_b$) do not
lead to useful results.  Also, when Eq.~\eqref{eq:cft-alg} is used in
applications, one wants the resulting integrals over the bosonic
variables to converge, and this requirement necessitates a non-compact
integration domain on the RHS of that
equation~\cite{SW,Efetov,VWZ,Zirnbauer:1996}.

Under the action of the subgroups SU($N_c$) and U($N_b,N_b$), the
operators $c_A^i$ and $\cba^i_A$ transform as follows,
\begin{alignat}{3}
  \label{eq:cd-trans-col}
  & g\in\text{SU}(N_c): & \qquad
  T_g \, c^i_A \, T_g^{-1}&=(g^{-1})^{ij} \, c^j_A \:, & \qquad 
  T_g \, \cba^i_A \, T_g^{-1}&= \cba^j_A \, g^{ji}\:, \\
  \label{eq:cd-trans-fl}
  & g\in\text{U}(N_b,N_b): & \qquad
  T_g \, c^i_A \, T_g^{-1}&=g_{AB}^{-1} \, c^i_B \:, & \qquad 
  T_g \, \cba^i_A \, T_g^{-1}&= \cba^i_B \, g_{BA}\:,
\end{alignat}
which can be shown using the Baker-Campbell-Hausdorff formula.

\subsection{Bose coherent states and projection onto the color-neutral
  sector} 
\label{sec:Bose}

We call a vector $|\mathcal N \rangle$ in the Fock space color-neutral
if it is invariant under SU($N_c$) transformations, i.e. $T_U|\mathcal
N \rangle=|\mathcal N \rangle$ for all $U\in\text{SU}(N_c)$.  The
subspace of Fock space spanned by these invariant vectors is called
the color-neutral subspace or sector.

The following argument closely parallels Ref.~\cite{BNSZ}.  With the
complex bosonic variables $\psi_a^i$, $\bar\psi_a^i$, $\varphi_a^i$,
and $\bar\varphi_a^i$ introduced in Sec.~\ref{sec:results}, Bose
coherent states are defined as
\begin{equation}
  |\Psi\rangle=\exp (\cba^i_a \psi ^i_a + \dba^i_a {\bar
  \varphi}^i_a)|0\rangle\:,\qquad 
  \langle\Psi|=\langle 0|\exp(\bar\psi ^i_a c^i_a 
  +\varphi^i_a d^i_a )\:. 
\end{equation}
They span the entire Fock space (or its dual).  Using
Eq.~\eqref{eq:cd-trans-col} we find
\begin{equation}
  \langle\Psi|T_U|\Psi\rangle =
    \exp\left(\bar{\psi}_{a}^{i}U^{ij}\psi_{a}^{j}+ 
    \bar{\varphi}_{a}^{i}U^{\dagger ij}\varphi_{a}^{j} \right)\:.
\end{equation}
The LHS of Eq.~\eqref{eq:cft-alg} can therefore be written
as
\begin{equation}
  \mathcal{Z} = \int_{{\rm SU}(N_c)} dU
  \exp\left(\bar{\psi}_{a}^{i}U^{ij}\psi_{a}^{j}+
    \bar{\varphi}_{a}^{i}U^{\dagger ij}\varphi_{a}^{j} \right) =
  \langle\Psi|P|\Psi\rangle\:, 
\end{equation}
where we have introduced the operator $P$ defined by 
\begin{equation}
  \label{eq:proj-col}
  P=\int_{\text{SU}(N_c)}dU\:T_U\:.
\end{equation}
This operator annihilates all states that are not color-neutral, while
leaving color-neutral states invariant (recall that the volume of
SU($N_c$) is unity).  Therefore, it is a projector onto the
color-neutral sector.  As advertised above, it is one possible
representation of such a projector, and we will now derive an
alternative form.

\subsection{Action of the flavor group in the color-neutral sector}

By definition, color-neutral vectors $|\mathcal N \rangle$ are
annihilated by all generators of sl($N_c$), i.e.  $\mathcal{E}^{ij} |
\mathcal{N}\rangle =0$ and $ \mathcal{H}^i |{\mathcal{N}}\rangle=0$.
Using Eq.~\eqref{eq:flsing} and the commutation relations
\eqref{eq:comm}, this requirement leads to
\begin{align} 
  \label{eq:cns}
  \left(\sum_{a=1}^{N_b}{\bar c}^i_a{c}^j_a-\sum_{a=1}^{N_b}{\bar
      d}^j_a{d}^i_a \right)|\mathcal N\rangle 
  =\delta^{ij}Q|\mathcal N\rangle  \:,
\end{align} 
where $Q$ is an integer.  Clearly, the color-neutral sector contains
the vacuum.  For $i=j$, the operator on the LHS of
Eq.~\eqref{eq:cns} counts the difference in the number of particles
and antiparticles for each color, the difference being equal to $Q$.

The color-neutral sector can be generated by acting on the vacuum
state with three types of operators,
\begin{align}
  \label{eq:type1a2ab}
  \text{type-1a:}\quad & \cba^{i}_{a}\dba^i_{b} \:,\nn\\
  \text{type-2a:}\quad & \epsilon_{i_1 \cdots i_{N_c}}{\bar
    c}^{i_1}_{a_1}\cba^{i_2}_{a_2}\cdots\cba^{i_{N_c}}_{a_{N_c}} \:,\\
  \text{type-2b:}\quad & \epsilon_{i_1 \cdots i_{N_c}}{\bar
    d}^{i_1}_{b_1}\dba^{i_2}_{b_2}\cdots\dba^{i_{N_c}}_{b_{N_c}}\:,\nn 
  \intertext{where $\epsilon$ denotes the totally antisymmetric
    Levi-Civita symbol which ensures that the resulting state is
    invariant under SU($N_c$) transformations.  In addition, the
    following types of operators make transformations in the
    color-neutral sector,}
  \text{type-1b:}\quad & c_a^id_b^i\:, \nn \\
  \label{eq:type1bcd}
  \text{type-1c:}\quad & \cba_a^ic_b^i\:,\\
  \text{type-1d:}\quad & d_a^i\dba_b^i\:. \nn
\end{align}
When acting on a color-neutral state, type-1 operators do not change
the $Q$-value of that state, whereas type-2a (type-2b) operators
increase (decrease) the $Q$-value by one.  Note, however, that for
$N_b<N_c$ the type-2 operators do not exist, which makes it
impossible to generate a vector in a non-zero $Q$-sector.  The range
of $Q$ is therefore given by
\begin{equation}
  \label{eq:rangeQ}
  Q=\left\{\begin{array}{cl}
      -\infty, \ldots, \infty & \text{ for } N_b\geq N_c \:, \\
      0 & \text{ for } N_b<N_c \:.
    \end{array} \right.
\end{equation}
(In the case of $N_b<N_c$, we are back to the bosonic color-flavor
transformation for the group $\mathrm
{U}(N_c)$~\cite{Zirnbauer:1997qm}. The Lie algebra of U($N_c$) has an
extra ${\mathrm U}(1)$ generator, and by requiring invariance under
${\mathrm U}(N_c)$, this ${\mathrm U}(1)$ generator eliminates all
non-zero $Q$-sectors.)

The action of the flavor group on the Fock space is defined by
Eq.~\eqref{eq:tg} with $g\in\text{U}(N_b,N_b)$.  We now choose the
color-neutral sector to be the carrier space of this representation.
The type-1 operators are the generators of the flavor group and do not
change the $Q$-value of a given state.  Therefore, under the action of
the flavor group, the color-neutral sector decomposes into invariant
subspaces labeled by $Q$, which we shall call ``$Q$-sectors''.  As
mentioned above, a $Q$-sector contains $Q$ more particles than
antiparticles for each color.

We now proof that the flavor group acts irreducibly in a given
$Q$-sector~\cite{Zirnbauer:1996}.  For this we need to show that from
any given state in the $Q$-sector we can reach any other state by the
action of the flavor group.  Equivalently, we can single out a
particular state $|\psi_Q\rangle$, defined by
\begin{equation}
  \label{eq:psiq}
  \begin{split}
    |\psi_{Q>0}\rangle &= (\epsilon_{i_1 \cdots i_{N_c}}{\bar
      c}^{i_1}_1\cba^{i_2}_2\cdots\cba^{i_{N_c}}_{N_c})^Q|0\rangle\:,\\
    |\psi_{Q=0}\rangle &= |0 \rangle \:,\\
    |\psi_{Q<0}\rangle &= (\epsilon_{i_1 \cdots i_{N_c}}{\bar
      d}^{i_1}_1\dba^{i_2}_2\cdots\dba^{i_{N_c}}_{N_c})^Q|0\rangle \:,
  \end{split}
\end{equation}
and show that (i) starting from this state, we can reach any other
state, and (ii) from that state we can return to $|\psi_Q\rangle$,
using type-1 operators only.

An arbitrary vector in a given $Q$-sector, which we should be able to
reach from $|\psi_Q\rangle$, is obtained by acting on the vacuum with
the appropriate number of type-1a operators and $Q$ more type-2a than
type-2b operators.  There are already $Q$ ($-Q$) unpaired type-2a
(type-2b) operators associated with $|\psi_Q\rangle$, so what remains
are pairs consisting of a type-2a and a type-2b operator.  Such a pair
can be expanded in terms of type-1a operators as
\begin{align}
  \epsilon_{i_1 \cdots i_{N_c}}{\bar c}^{i_1}_{a_1}\cba^{i_2}_{a_2}
  \cdots\cba^{i_{N_c}}_{a_{N_c}}\epsilon_{j_1 \cdots j_{N_c}}{\bar
    d}^{j_1}_{b_1}\dba^{j_2}_{b_2}\cdots\dba^{j_{N_c}}_{b_{N_c}}
  =\sum_{\sigma\in S_{N_c}}{\text {sgn}}(\sigma)\cba^{i_1}_{a_1}
  \dba_{\sigma(b_1)}^{i_1}\cdots \cba^{i_{N_c}}_{a_{N_c}}
  \dba_{\sigma(b_{N_c})}^{i_{N_c}}
\end{align}
and is therefore in the algebra of the flavor group.  The type-1c and
type-1d operators obey the commutation relations
\begin{align}
  [\cba^i_ac^i_{b_1}, \epsilon_{i_1 \cdots i_{N_c}}{\bar
    c}^{i_1}_{b_1}\cba^{i_2}_{b_2}\cdots\cba^{i_{N_c}}_{b_{N_c}}]
  &=\epsilon_{i_1 \cdots i_{N_c}}{\bar
    c}^{i_1}_{a}\cba^{i_2}_{b_2}\cdots\cba^{i_{N_c}}_{b_{N_c}}\:, \nn\\
  [d^i_{b_1}\dba^i_a, \epsilon_{i_1 \cdots i_{N_c}}{\bar
    d}^{i_1}_{b_1}\dba^{i_2}_{b_2}\cdots\dba^{i_{N_c}}_{b_{N_c}}]
  &=\epsilon_{i_1 \cdots i_{N_c}}{\bar
    d}^{i_1}_{a}\dba^{i_2}_{b_2}\cdots\dba^{i_{N_c}}_{b_{N_c}}\:.
\end{align} 
Thus, they enable us to change the flavor indices of the type-2
operators.  We can thus reach any state in a given $Q$-sector by
acting with the corresponding number and subtypes of type-1 operators
on $|\psi_Q\rangle$.  Furthermore, type-1a operators can be undone by
type-1b operators, while type-1c and type-1d operators can undo
themselves.  We can therefore go from an arbitrary state in the
$Q$-sector back to $|\psi_Q\rangle$.  In other words, $|\psi_Q\rangle$
is a cyclic vector in the $Q$-sector under the action of the flavor
group, which implies irreducibility.

\subsection{Generalized coherent states and projection onto the
  color-neutral sector} 

Generalized coherent states are described in detail in
Ref.~\cite{Perelomov:1986}.  They are useful for our purposes because
they allow a resolution of the identity operator.  For a Lie group $G$
and an irreducible representation $T_g$, a set of generalized coherent
states is obtained by acting on a state $|\psi_T\rangle$ in the
carrier space of $T_g$ with all elements of $T_g$.  This results in
the set $\{T_g|\psi_T\rangle\}$ which, in general, is overcomplete.
If $H$ is the maximal subgroup of $G$ such that
$T_h|\psi_T\rangle\propto|\psi_T\rangle$ for all $h\in H$, the
subgroup $H$ is called the isotropy subgroup of $|\psi_T\rangle$, and
the set of generalized coherent states can be parameterized without
overcounting by the elements of the coset space $G/H$.  (If the
subgroup $H$ is not maximal, some overcounting remains.)

We now set $G=\text{U}(N_b,N_b)$ and consider the representation $T_g$
of Eq.~\eqref{eq:tg} (with $g\in G$) which acts irreducibly in a given
$Q$-sector.  For the starting vector we choose the vector
$|\psi_Q\rangle$ defined in Eq.~\eqref{eq:psiq}, resulting in the
(overcomplete) set of generalized coherent states
$\{T_g|\psi_Q\rangle\}$.  The identity operator in this $Q$-sector is
then given by~\cite{Perelomov:1986}
\begin{equation}
  \label{eq:projQ}
  {\1}_{Q}=C_{Q}\int_G dg \ T_g|\psi_{Q}\rangle
  \langle\psi_{Q}|T_{g}^{-1}\:, 
\end{equation}
where $dg$ is the invariant measure of U($N_b,N_b$) and $C_Q$ is a
normalization factor defined by
\begin{equation}
  \label{eq:cqm}
  C_Q^{-1}=\frac{1}{N_Q}\int_G dg \ \langle \psi_{Q}|T_g|\psi_{Q}
  \rangle \langle \psi_{Q}|T_{g}^{-1} | \psi_{Q} \rangle\:.
\end{equation}
Here, $N_Q$ is the norm of $|\psi_Q\rangle$,
\begin{equation}
  \label{eq:nq}
  N_Q=\langle \psi_Q | \psi_Q\rangle
  =\prod_{n=0}^{N_c-1}\frac{(|Q|+n)!}{n!}\:. 
\end{equation}
A detailed calculation of $N_Q$ is given in App.~\ref{app:norm}. 

The operator in Eq.~\eqref{eq:projQ} annihilates all states that are
not color-neutral, as well as color-neutral states corresponding to a
different value of $Q$.  Thus, it is a projector onto the $Q$-sector.
We can therefore write the projector onto the color-neutral sector as
\begin{equation}
  \label{eq:proj}
  P=\bigoplus_Q \1_Q \:,
\end{equation}
where the sum runs over the values of $Q$ given in
Eq.~\eqref{eq:rangeQ}.  Identifying this projection operator with the
one in Eq.~\eqref{eq:proj-col} yields
\begin{align}
  \label{eq:ZQ}
  \mathcal{Z}&=\sum_Q\mathcal{Z}_Q \qquad\text{with} \nn\\
  \mathcal{Z}_Q &= \langle\Psi|\1_Q|\Psi\rangle = 
  \langle 0|\exp(\bar{\psi}^i_ac^i_a + {\varphi}^i_ad^i_a) \1_Q 
  \exp (\cba^i_a \psi^i_a + \dba^i_a {\bar \varphi}^i_a)|0\rangle \:.
\end{align}

\subsection{Parameterization of the coherent states}

The maximal compact subgroup of the flavor group $G=\text{U}(N_b,N_b)$
is $H=H_+\times H_-=\mathrm U(N_b)\times \mathrm U(N_b)$ with elements
$h=\diag(h_+,h_-)$, where $h_\pm\in\text{U}(N_b)$. The corresponding
Fock operators are
\begin{equation}
  \label{eq:th}
  T_{h}=\exp\left[{\cba}^i_a(\ln h_+)_{ab}c^i_b
  -d^i_a(\ln h_-)_{ab}\dba^i_b\right] \:.
\end{equation}
For $Q=0$, these operators stabilize the vacuum,
\begin{equation}
  \label{eq:stab}
  T_h|0\rangle=\exp\left(-N_c\tr\ln
  h_-\right)|0\rangle=({\det}^{-N_c}h_-)\:|0\rangle\:, 
\end{equation}
and therefore the set of coherent states for $Q=0$ can be
parameterized without overcounting by the elements of the non-compact
coset space $S=G/H=\text{U}(N_b,N_b)/[\text{U}(N_b)\times U(N_b)]$.  (We
will use the same coset space also for $Q\neq0$, see
Sec.~\ref{sec:ZQ}.)

To arrive at this parameterization, we first use the canonical
projection $\pi:G\mapsto G/H$ which assigns to each $g\in G$ the
corresponding equivalence class $gH$.  We then choose a representative
group element $s(\pi(g))$ from each equivalence class and write an
arbitrary group element $g$ as the product $g=s(\pi(g))h(g)$.  The
coset element $s(\pi(g))$ can be parameterized using projective
coordinates $Z$, see Eqs.~(5.8n) and (5.28) in Ch.~9 of
Ref.~\cite{gilmore},
\begin{align}
  \label{eq:ts}
  s(\pi(g))&\equiv s(Z,\Zd)=
  \begin{pmatrix}
    (1-Z\Zd)^{-1/2} & Z(1-\Zd Z)^{-1/2} \\
    \Zd(1-Z\Zd)^{-1/2} & (1-\Zd Z)^{-1/2}
  \end{pmatrix} \nn \\
  &= \begin{pmatrix} 1 & Z \\ 0 & 1 \end{pmatrix}
  \begin{pmatrix} 
    (1-Z\Zd)^{1/2} & 0 \\ 
    0 & (1-\Zd Z)^{-1/2} 
  \end{pmatrix}
  \begin{pmatrix} 1 & 0 \\ \Zd & 1 \end{pmatrix}\:.
\end{align} 
Here, $Z$ is an $N_b\times N_b$ complex matrix with the constraint
$|ZZ^{\dag}| \leq 1$.  We have $s=s^\dagger$ and
$s^{-1}=s(-Z,-Z^{\dag})$.  Also, $s(Z,Z^\dagger)$ satisfies the
pseudo-unitarity condition
$s\diag(\1_{N_b},-\1_{N_b})s^\dagger=\diag(\1_{N_b},-\1_{N_b})$.
Using the decomposition \eqref{eq:ts}, the Fock space operator
corresponding to $s(Z,Z^\dagger)$ becomes
\begin{align} 
  T_{s(Z,Z^\dagger)}&=\exp(\cba^i_{a}Z_{ab}\dba^i_b)
  \exp\left[\frac12\cba_a^i \ln(1-Z\Zd)_{ab}c^i_b
   +\frac12d^i_a\ln(1-\Zd Z)_{ab}\dba^i_b\right]\nn\\
  & \quad \times \exp(-d^i_a\Zd_{ab}c^i_b)\:.
\end{align} 
The coset space $G/H$ has a $G$-invariant measure \cite{hua} that has
already been given in Eq.~\eqref{eq:dz}.  We can therefore use
Eq.~\eqref{eq:homo} to rewrite the integral \eqref{eq:projQ} over $G$
as an integral over $H$ and $S=G/H$,
\begin{align} 
  \label{eq:proj-int}
  {\1}_{Q}&=C_{Q}\int_G dg \ T_g|\psi_{Q}\rangle \langle
  \psi_{Q}|T_{g}^{-1}\nn \\ 
  &= C_Q \int_{G/H} d \mu(Z,Z^\dagger)\int_H dh \ 
  T_{s}T_h|\psi_{Q}\rangle \langle \psi_{Q}|T_h^{-1}T_{s}^{-1}\:.
\end{align} 
The Haar measure $dh$ of $H$ is normalized to unity.

\subsection{Calculation of $\mathcal{Z}_0$}
\label{sec:Z0}

For $Q=0$, Eq.~\eqref{eq:stab} tells us that the integration over $H$
in Eq.~\eqref{eq:proj-int} is trivial, and we are left with
\begin{equation}
  \mathcal{Z}_0=C_0\!\!\int\limits_{G/H}\!\!d\mu(Z,Z^{\dag}) 
  \langle 0|\exp\left(\bar{\psi}^i_ac^i_a + {\varphi}^i_ad^i_a\right)
  T_s|0\rangle \langle0|T_s^{-1} \exp \left(\cba^i_a \psi^i_a +
  \dba^i_a {\bar \varphi}^i_a\right)|0\rangle\:.  
\end{equation}
Defining the notation
$|Z\rangle=\exp(\cba^i_{a}Z_{ab}\dba^i_{b})|0\rangle$ and $\langle
Z|=\langle0|\exp(d^i_aZ^\dagger_{ab}c^i_b)$, we find
\begin{equation}
  T_s|0\rangle={\det}^{N_c/2}(1-Z\Zd)|Z\rangle 
\end{equation}
and
\begin{align}
  \langle 0|&\exp\left(\bar{\psi}^i_ac^i_a + {\varphi}^i_ad^i_a\right)
  T_s|0\rangle \langle0|T_s^{-1} \exp \left(\cba^i_a \psi^i_a +
  \dba^i_a {\bar \varphi}^i_a\right)|0\rangle \nn\\ 
  &= {\det}^{N_c}(1-ZZ^\dagger ) \langle 0|\exp\left(\bar\psi^i_ac^i_a
  + {\varphi}^i_ad^i_a\right)|Z\rangle \langle Z|\exp \left(\cba^i_a
  \psi^i_a + \dba^i_a {\bar \varphi }^i_a\right)|0\rangle \nn \\ 
  &= {\det}^{N_c}(1-ZZ^\dagger) \exp \left(\bar\psi^i_aZ_{ab}
    \varphi^i_b+\bar\varphi^i_a\Zd_{ab}\psi^i_b\right)\:. 
\end{align} 
Thus
\begin{equation}
  \label{eq:q0}
  \mathcal{Z}_0=C_0 \int_{|Z\Zd|\leq 1}D(Z,Z^\dagger) \exp
  \left(\bar\psi^i_aZ_{ab}\varphi^i_b+\bar\varphi^i_a\Zd_{ab}
    \psi^i_b\right) 
\end{equation}
with
\begin{equation}
  \label{eq:Dz}
  D(Z,\Zd)=\frac{dZd\Zd}{{\det}^{2N_b-N_c}(1-Z\Zd)}\:.
\end{equation}
Not surprisingly, this is the same result as in
Ref.~\cite{Zirnbauer:1997qm} for the color group U($N_c$).  From
Eq.~\eqref{eq:cqm} we have
\begin{align}
  C_0^{-1}=\frac{1}{N_0}\int_{G} dg \langle 0|T_g|0\rangle
  \langle0|T_g^{-1} |0\rangle 
  =\int_{|Z^{\dagger} Z| \leq 1} D(Z,Z^{\dagger})\:.
\end{align}
An explicit calculation of $C_0$ is given in App.~\ref{app:rad}.

Note that for $2N_b>N_c$, $D(Z,Z^{\dagger})$ becomes divergent at the
boundary of the integration domain. This divergence is due to the
non-compactness of the symmetric space ${\mathrm U}(N_b,N_b)/[{\mathrm
  U}(N_b)\times {\mathrm U}(N_b)]$ and is a feature of the bosonic
color-flavor transformation. In Ref.~\cite{Zirnbauer:1996}, this
divergence is canceled by the measure of the fermionic degrees of
freedom, and the non-compact supersymmetric coset space has a flat
measure if there is an equal number of bosons and fermions.  In the
fermionic case~\cite{fermi-cft}, the integral on the RHS is over the
compact symmetric space $\mathrm U(2N_f)/[{\mathrm U(N_f)\times
  \mathrm U(N_f)}]$, and there is no divergence problem.

Note also that for $2N_b>N_c$, there are divergences in both numerator
and denominator of the above formula.  Apart from $D(Z,Z^\dagger)$,
the integrands are analytic on the entire coset space, therefore the
divergences in numerator and denominator are of the same degree and
the ratio must be finite.  We will show this for a simple example in
App.~\ref{app:ex1-div}.  However, in the general case it is not
obvious how to cancel the divergences, and even if it were possible,
the resulting expressions might not be simple enough to be useful in
applications.  That is why in Sec.~\ref{sec:char} we will use another
method to extend the range of $N_b$ in which all terms in the
transformation are finite.

\subsection{Calculation of $\mathcal{Z}_Q$ for $Q\neq0$}
\label{sec:ZQ}

\begin{figure}[tb]
  \begin{center}
    \unitlength1cm
    \begin{picture}(7.1,4)
      \put(0.0,1.9){\makebox(0,0){${\Gamma}\::$}}
      \put(0.5,1.1){$\begin{array}{c} 
          \underbrace{
            \begin{picture}(3.5,0.5)(0,-0.15)
              \put(0.0,0.0){\line(1,0){0.5}} \put(0.0,0.5){\line(1,0){0.5}}
              \put(0.0,0.0){\line(0,1){0.5}} \put(0.5,0.0){\line(0,1){0.5}}
              \put(0.75,0.25){\makebox(0,0){$\otimes$}}
              \put(1.0,0.0){\line(1,0){0.5}} \put(1.0,0.5){\line(1,0){0.5}}
              \put(1.0,0.0){\line(0,1){0.5}} \put(1.5,0.0){\line(0,1){0.5}}
              \put(1.75,0.25){\makebox(0,0){$\otimes$}}
              \put(2.25,0.25){\makebox(0,0){$\cdots$}}
              \put(2.75,0.25){\makebox(0,0){$\otimes$}}
              \put(3.0,0.0){\line(1,0){0.5}} \put(3.0,0.5){\line(1,0){0.5}}
              \put(3.0,0.0){\line(0,1){0.5}} \put(3.5,0.0){\line(0,1){0.5}}
            \end{picture}}\\
          N_c\cdot Q
        \end{array}$}
      \put(4.5,1.8){\vector(1,0){1.25}}
      \put(6.5,1.85){\makebox(0,0){$\hat\Gamma\::$}}
    \end{picture}
    \begin{picture}(5,4)
      \put(3.0,4.0){\vector(1,0){2}} \put(2.5,4.0){\vector(-1,0){2}}
      \put(2.75,4.0){\makebox(0,0){$Q$}}
      \put(0.0,2.0){\vector(0,1){1.5}} \put(0.0,1.5){\vector(0,-1){1.5}}
      \put(0.0,1.75){\makebox(0,0){$N_c$}}
      \put(0.5,0.0){\line(1,0){4.5}} \put(0.5,0.5){\line(1,0){4.5}}
      \put(0.5,1.0){\line(1,0){4.5}} \put(0.5,1.5){\line(1,0){4.5}}
      \put(0.5,2.0){\line(1,0){4.5}} \put(0.5,2.5){\line(1,0){4.5}}
      \put(0.5,3.0){\line(1,0){4.5}} \put(0.5,3.5){\line(1,0){4.5}}
      \put(0.5,0.0){\line(0,1){3.5}} \put(1.0,0.0){\line(0,1){3.5}}
      \put(1.5,0.0){\line(0,1){3.5}} \put(2.0,0.0){\line(0,1){3.5}}
      \put(2.5,0.0){\line(0,1){3.5}} \put(3.0,0.0){\line(0,1){3.5}}
      \put(3.5,0.0){\line(0,1){3.5}} \put(4.0,0.0){\line(0,1){3.5}}
      \put(4.5,0.0){\line(0,1){3.5}} \put(5.0,0.0){\line(0,1){3.5}}
    \end{picture}
    \caption{The (reducible) product of fundamental representations of
      U($N_b$) contains an irreducible representation of U($N_b$) with
      symmetric color indices and antisymmetric flavor indices.  Here
      $Q>0$.}
    \label{fig:u-n-young-graph}
  \end{center}
\end{figure}
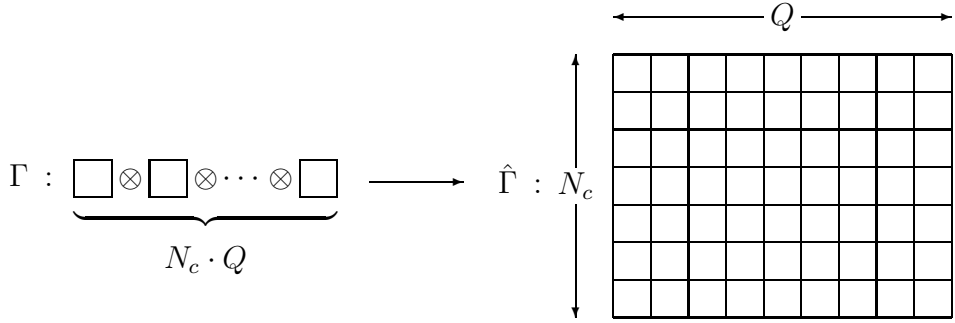

Let us start with the case of $Q>0$; the case of $Q<0$ follows
analogously.  Similar to Ref.~\cite{fermi-cft}, the idea is to relate
the state $|\psi_Q\rangle$ to $|0\rangle$ by the action of the
creation operators.  Starting from the integrand of Eq.~\eqref{eq:ZQ},
we perform the following manipulations,
\begin{align} 
\label{eq:long}
  &\langle0|\exp(\bar\psi^i_ac^i_a + {\varphi}^i_ad^i_a)T_g
  |\psi_{Q}\rangle \langle\psi_{Q}|T_{g}^{-1} \exp (\cba^i_a\psi^i_a +
  \dba^i_a {\bar \varphi}^i_a)|0\rangle \nn \\ 
  &=\langle 0|\exp(\bar\psi^i_ac^i_a +
  {\varphi}^i_ad^i_a)(\epsilon_{i_1 \cdots i_{N_c}} 
  {T_g\bar c}^{i_1}_1T_g^{-1}T_g\cba^{i_2}_2T_g^{-1}\cdots
  T_g\cba^{i_{N_c}}_{N_c}T^{-1}_g)^Q T_g|0\rangle \nn \\
  &\quad\times \langle 0|T_g^{-1}(\epsilon_{i_1 \cdots i_{N_c}}{T_g
  c}^{i_1}_1 T_g^{-1}T_g c^{i_2}_2T_g^{-1}\cdots T_g
  c^{i_{N_c}}_{N_c}T_g^{-1})^Q \exp (\cba^i_a \psi ^i_a +\dba^i_a
  {\bar \varphi}^i_a)|0\rangle \nn \\ 
  &={\det}^{N_c}(1-ZZ^\dagger ) \exp(\bar\psi^i_aZ_{ab}\varphi^i_b+
  \bar\varphi^i_a\Zd_{ab}\psi^i_b) \nn \\ 
  &\quad\times\left[(\epsilon_{i_1\cdots i_{N_c}}
    {\hat{\bar\psi}^{i_1}_{a^1_1}} \cdots {\hat{\bar
        \psi}}^{i_{N_c}}_{a^1_{N_c}})\cdots (\epsilon_{j_1 \cdots
      j_{N_c}}{\hat {\bar \psi}^{j_1}_{a^Q_1}}\cdots {\hat{\bar
        \psi}}^{j_{N_c}}_{a^Q_{N_c}})\right]  \Gamma^{(a^1_1 \cdots
    a^1_{N_c})\cdots (a^Q_1 \cdots  a^Q_{N_c})}_{(1\cdots
    N_c)\cdots(1\cdots N_c)} \nn \\ 
  &\quad\times \left[(\epsilon_{i'_1 \cdots i'_{N_c}}{\hat
      {\psi}^{i'_1}_{b^1_1}} \cdots {\hat{
        \psi}}^{i'_{N_c}}_{b^1_{N_c}})\cdots (\epsilon_{j'_1 \cdots
      j'_{N_c}}{\hat {\psi}^{j'_1}_{b^Q_1}}\cdots 
    {\hat{\psi}}^{j'_{N_c}}_{b^Q_{N_c}})\right]
  {\bar\Gamma}_{(1\cdots N_c)\cdots(1\cdots N_c)}^{(b^1_1 \cdots
    b^1_{N_c}) \cdots (b^Q_1 \cdots b^Q_{N_c})}\:.
\end{align}
In the first step, we have used Eq.~\eqref{eq:psiq} and inserted
$T_g^{-1}T_g$ between each pair of creation and annihilation
operators.  In the second step, which involves a tedious but
straightforward calculation, we have used the transformation
properties of the Fock space operators given in
Eq.~\eqref{eq:cd-trans-fl}, the Baker-Campbell-Hausdorff formula, and
the coset decomposition of the group elements, $g=s(Z,\Zd)h(g)$.  We
have also defined
\begin{equation}
  {\hat {\bar \psi}}^{i}_{a}= {\bar \psi}^i_b
  (1-Z\Zd)^{\frac{1}{2}}_{ba} \:,\qquad
  {\hat {\psi}}^{i}_{a}= (1-Z\Zd)^{\frac{1}{2}}_{ab}\psi^i_b
\end{equation}
and
\begin{align} 
  \Gamma^{(a^1_1 \cdots a^1_{N_c})\cdots (a^Q_1 \cdots
    a^Q_{N_c})}_{(b^1_1 \cdots b^1_{N_c})\cdots(b^Q_1 \cdots b^Q_{N_c})}
  =({h_+}_{a^1_1b^1_1} \cdots {h_+}_{a^1_{N_c}b^1_{N_c}})\cdots 
  ({h_+}_{a^Q_1b^Q_1} \cdots {h_+}_{a^Q_{N_c}b^Q_{N_c}}) \:.
\end{align}
The hypermatrix $\Gamma$ represents the direct product of $N_c\cdot Q$
fundamental representations of $H_+=\text{U}(N_b)$, see
Fig.~\ref{fig:u-n-young-graph}.  Note that $H_-$ does not appear here.
Inserting Eq.~\eqref{eq:long} into Eq.~\eqref{eq:ZQ}, we obtain
\begin{align} 
  \label{eq:ZQint}
  \mathcal{Z}_Q&=C_Q\int_{G/H}d \mu(Z,Z^\dagger)
  \int_H dh \langle0|\exp(\bar\psi^i_ac^i_a + {\varphi}^i_ad^i_a) 
  T_{s(Z,Z^{\dagger})}T_h|\psi_{Q}\rangle \nn\\
  & \hspace*{49mm}\times \langle\psi_{Q}|T_h^{-1}T_{s(Z, Z^\dagger)}^{-1} 
  \exp (\cba^i_a \psi ^i_a + \dba^i_a {\bar \varphi}^i_a)|0\rangle \nn \\
  &= \int_{|Z\Zd|\leq 1}D(Z,Z^{\dag})\:\exp\left(\bar{\psi}_a^i
    Z_{ab}\varphi_b^i +\bar{\varphi}_a^i\Zd_{ab}\psi_b^i \right)\:\chi_{Q}\:,
\end{align} 
where we have defined 
\begin{align}
  \label{eq:chiq+}
  \chi_{Q} &= C_Q\left\{(\epsilon_{i_1 \cdots i_{N_c}}{\hat {\bar
        \psi}^{i_1}_{a^1_1}} \cdots {\hat{\bar
        \psi}}^{i_{N_c}}_{a^1_{N_c}})\cdots (\epsilon_{j_1 \cdots
      j_{N_c}}{\hat {\bar \psi}^{j_1}_{a^Q_1}}\cdots 
    {\hat{\bar \psi}}^{j_{N_c}}_{a^Q_{N_c}})\right\}\nn\\
  &\quad\times\left\{(\epsilon_{i'_1 \cdots i'_{N_c}}{\hat
      {\psi}^{i'_1}_{b^1_1}} \cdots {\hat{
        \psi}}^{i'_{N_c}}_{b^1_{N_c}})\cdots (\epsilon_{j'_1 \cdots
      j'_{N_c}}{\hat {\psi}^{j'_1}_{b^Q_1}}\cdots  
    {\hat{\psi}}^{j'_{N_c}}_{b^Q_{N_c}})\right\} \nn\\
  &\quad\times \int_{{\mathrm U}(N_b)} dh_+\
  \left[{\Gamma}^{(a^1_1 \cdots a^1_{N_c})\cdots (a^Q_1 \cdots 
    a^Q_{N_c})}_{(1 \cdots N_c) \cdots (1 \cdots
        N_c)}{\bar\Gamma}_{(1\cdots N_c) \cdots(1\cdots N_c)}^{(b^1_1
        \cdots b^1_{N_c})\cdots (b^Q_1 \cdots b^Q_{N_c})}\right]\:.
\end{align}
It follows from the definition of $\Gamma$ that the term in square
brackets is totally symmetric under the exchange of $a^i_a$ with
$a^{i'}_a$ and of $b^i_a$ with $b^{i'}_a$.  Because of the contraction
with the totally antisymmetric tensor $\epsilon$, the terms in curly
brackets are totally antisymmetric under the exchange of $a^i_a$ with
$a^i_{a'}$ and of $b^i_a$ with $b^i_{a'}$.  Therefore, after the
contractions of the $a^i_a$'s and $b^i_a$'s, only terms with the
correct symmetry properties survive, i.e. symmetric in color and
antisymmetric in flavor.  In other words, when the (reducible)
direct-product representation $\Gamma$ is decomposed into irreducible
representations, only the irreducible representation $\hat \Gamma$
shown in Fig.~\ref{fig:u-n-young-graph} contributes to $\chi_Q$.

This observation enables us to perform the integration over $H_+$ in
the same way as in Ref.~\cite{fermi-cft}.  We use the group theoretic
result that for irreducible unitary representations $\Gamma^{r}$ and
$\Gamma^{r'}$ of a compact Lie group $G$,
\begin{equation}
\label{eq:ortho_rel}
  \int_{G}dg\:\bar\Gamma^r_{ab}\Gamma^{r'}_{a'b'}
  =\frac{1}{d_r}\delta_{aa'}\delta_{bb'}\delta^{rr'}\int_{G}dg\:, 
\end{equation}
where $d_r$ is the dimension of $\Gamma^r$.  The group volume of
U($N_b$) is normalized to unity as mentioned above.

Comparing Eq.~\eqref{eq:chiq+} with Eq.~(41) of Ref.~\cite{fermi-cft},
we realize that we can use (a slightly modified version of) the result
in Eq.~(46) of that reference and thus obtain
\begin{align}
  \label{eq:chiq}
  \chi_Q=\frac{(N_c!)^Q}{d^{N_b}_{N_c,Q}(Q!)^{N_c}}\:C_Q\:
  {\det}^Q \mathcal M \:,
\end{align} 
where we have defined the $N_c\times N_c$ matrix $\mathcal M$ by
${\mathcal M}^{ij}={\bar \psi}^i_a(1-Z\Zd)_{ab}\psi^j_b$.  The symbol
$d^{N_b}_{N_c,Q}$ denotes the dimension of the irreducible
representation of $\mathrm U(N_b)$ specified by a Young diagram with
$N_c$ rows and $Q$ columns, i.e.\ of the representation $\hat\Gamma$
in Fig.~\ref{fig:u-n-young-graph}.  (For $N_b\le N_c$, this dimension
is equal to one.)  We will see in a moment that the prefactor of
$C_Q{\det}^Q\mathcal M$ in Eq.~\eqref{eq:chiq} is in fact irrelevant.

We now use Eq.~\eqref{eq:cqm} to calculate the normalization factor
$C_Q$.  Using similar methods as in the calculation of $\chi_Q$, we
obtain
\begin{align}
  &\langle \psi_{Q}| \ T_g|\psi_{Q} \rangle \nn\\
  &={\det}^{\frac{N_c}{2}}(1-ZZ^{\dagger}) \langle \psi_Q |
  \left\{(\epsilon_{i_1 \cdots i_{N_c}}{ {\cba}^{i_1}_{c^1_1}} \cdots
    {\cba}^{i_{N_c}}_{c^1_{N_c}}) \cdots (\epsilon_{j_1 \cdots j_{N_c}}
    {{\cba}^{j_1}_{c^Q_1}}\cdots
    {{\cba}}^{j_{N_c}}_{c^Q_{N_c}})\right\}|0\rangle  \nn\\
  &\quad\times\left[ (1\!-\!Z\Zd)^{\frac{1}{2}}_{c^1_1a^1_1}\cdots
    (1\!-\!Z\Zd)^{\frac{1}{2}}_{c^1_{N_c}a^1_{N_c}}\right] \cdots
  \left[ (1\!-\!Z\Zd)^{\frac{1}{2}}_{c^Q_1a^Q_1}\cdots
    (1\!-\!Z\Zd)^{\frac{1}{2}}_{c^Q_{N_c}a^Q_{N_c}}\right] \nn\\
  &\quad\times\left[ {h_+}_{a^1_11} \cdots {h_+}_{a^1_{N_c}N_c}\right]
  \cdots \left[{h_+}_{a^Q_11} \cdots {h_+}_{a^Q_{N_c}N_c}\right] \nn\\
  &= N_Q\, {\det}^{\frac{N_c}{2}}(1-ZZ^{\dagger})\,
  {\Gamma}^{(a^1_1 \cdots a^1_{N_c})\cdots (a^Q_1 \cdots
    a^Q_{N_c})}_{(1 \cdots N_c) \cdots (1 \cdots N_c)} \nn\\
  &\quad\times \sum_{\sigma_1 \cdots \sigma_Q}\left[{\rm sgn}(\sigma_1)
    (1-Z\Zd)^{\frac{1}{2}}_{\sigma_1(1)a^1_1}\cdots
    (1-Z\Zd)^{\frac{1}{2}}_{\sigma_1({N_c})a^1_{N_c}} \right]\nn\\
  &\hspace*{20mm} \cdots\left[{\rm sgn}(\sigma_Q)
    (1-Z\Zd)^{\frac{1}{2}}_{\sigma_Q(1)a^Q_1}\cdots
    (1-Z\Zd)^{\frac{1}{2}}_{\sigma_Q({N_c})a^Q_{N_c}} \right] \:,
\end{align}
where we have used $N_Q=\langle \psi_Q|\psi_Q\rangle$.  The symbol
$\sigma$ and the symbol $\rho$ in the equation below denote elements
of the permutation group $S_{N_c}$.  Analogously, we find
\begin{align} 
  \langle \psi_{Q}| \ T_g^{-1}|\psi_{Q}\rangle 
  &= N_Q \, {\det}^{\frac{N_c}{2}}(1-ZZ^{\dagger}) \,
  {\bar\Gamma}_{(1\cdots N_c) \cdots(1\cdots N_c)}^{(b^1_1
    \cdots b^1_{N_c})\cdots (b^Q_1 \cdots b^Q_{N_c})} \\
  &\quad\times \sum_{\rho_1\cdots \rho_Q}\left[{\rm sgn}(\rho_1)
    (1-Z\Zd)^{\frac{1}{2}}_{b^1_1\rho_1(1)}\cdots
    (1-Z\Zd)^{\frac{1}{2}}_{b^1_{N_c}\rho_1({N_c})}\right]\nn \\
  &\hspace*{19mm} \cdots\left[{\rm sgn}(\rho_Q)
    (1-Z\Zd)^{\frac{1}{2}}_{b^Q_1\rho_Q(1)}\cdots
    (1-Z\Zd)^{\frac{1}{2}}_{b^Q_{N_c}\rho_Q({N_c})}\right]\:.\nn
\end{align}
Combining these two results, we have from Eq.~\eqref{eq:cqm}
\begin{align} 
    C_Q^{-1}&=\frac{1}{N_Q}\int_{{\mathrm U}(N_b,N_b)} dg
  \langle\psi_{Q}| \ T_g|\psi_{Q} \rangle \langle \psi_{Q}|T_{g}^{-1}
  | \psi_{Q} \rangle \nn \\ 
  &=N_Q \int_{|Z\Zd|\leq 1} d\mu(Z,Z^{\dagger})\ 
  {\det}^{{N_c}}(1-ZZ^{\dagger})\nn \\ 
  &\quad\times  \sum_{\sigma_1 \cdots \sigma_Q}\left[{\rm sgn}(\sigma_1)
    (1-Z\Zd)^{\frac{1}{2}}_{\sigma_1(1)a^1_1}\cdots
    (1-Z\Zd)^{\frac{1}{2}}_{\sigma_1({N_c})a^1_{N_c}}\right]\nn\\
  &\hspace{20mm}\cdots \left[{\rm sgn}(\sigma_Q)
    (1-Z\Zd)^{\frac{1}{2}}_{\sigma_Q(1)a^Q_1}\cdots
    (1-Z\Zd)^{\frac{1}{2}}_{\sigma_Q({N_c})a^Q_{N_c}}\right]\nn\\
  &\quad\times \sum_{\rho_1 \cdots \rho_Q}\left[{\rm sgn}(\rho_1)
    (1-Z\Zd)^{\frac{1}{2}}_{b^1_1\rho_1(1)}\cdots
    (1-Z\Zd)^{\frac{1}{2}}_{b^1_{N_c}\rho_1({N_c})}\right]\nn\\
  &\hspace{20mm}\cdots \left[{\rm sgn}(\rho_Q)
    (1-Z\Zd)^{\frac{1}{2}}_{b^Q_1\rho_Q(1)}\cdots
    (1-Z\Zd)^{\frac{1}{2}}_{b^Q_{N_c}\rho_Q({N_c})}\right]\nn\\
  &\quad\times\int_{{\mathrm U}(N_b)} dh_+\ \Gamma^{(a^1_1 \cdots
    a^1_{N_c})\cdots (a^Q_1 \cdots a^Q_{N_c})}_{(1\cdots N_c)\cdots(1\cdots
    N_c)}{\bar\Gamma}_{(1\cdots N_c)\cdots(1\cdots N_c)}^{(b^1_1 \cdots
    b^1_{N_c})\cdots (b^Q_1 \cdots b^Q_{N_c})} \:.
\end{align}
The integration over $\mathrm U(N_b)$ is of the same type as in the
calculation of $\chi_Q$, and using the same method we obtain
\begin{align}
  \label{eq:norm}
  C_Q^{-1}=\frac{N_Q(N_c!)^Q}{d^{N_b}_{N_c,Q}(Q!)^{N_c}}
  \int_{|Z\Zd|\leq 1} D(Z, Z^{\dagger})\ {\det}^Q(1-Z\Zd)_{[N_c]} \:,
\end{align}
where $(1-Z\Zd)_{[N_c]}$ denotes the upper left $N_c\times N_c $ block
of the $N_b\times N_b$ matrix $(1-Z\Zd)$.  Recall that non-zero
$Q$-sectors only exist for $N_b\geq N_c$, so this notation always
makes sense.

We now combine Eqs.~\eqref{eq:ZQint}, \eqref{eq:chiq}, and
\eqref{eq:norm} to obtain for $Q>0$
\begin{align}
  \label{eq:q+}
  \mathcal{Z}_Q={\mathcal C}_Q{\int_{|Z\Zd|\leq 1}D(Z,Z^{\dag})\: 
    \exp\left(\bar{\psi}_a^i Z_{ab}\varphi_b^i
      +\bar{\varphi}_a^i\Zd_{ab}\psi_b^i \right)\: 
    {\det}^Q{\mathcal M}}\:, 
\end{align} 
where we have defined
\begin{equation}
  \label{eq:radial-int}
  {\mathcal C}_Q^{-1}={N_Q} \int_{|Z\Zd|\leq 1}D(Z, Z^{\dagger})\ 
  {\det}^Q(1-Z\Zd)_{[N_c]}\:.
\end{equation}
The explicit calculation of this integral is performed in
App.~\ref{app:rad}.  As anticipated, the nontrivial prefactors in
Eqs.~\eqref{eq:chiq} and \eqref{eq:norm} have dropped out.

For $Q<0$, the calculation proceeds in exact analogy, and we obtain
\begin{align}
  \label{eq:q-}
  \mathcal{Z}_Q=\mathcal{C}_{|Q|}\int_{|Z\Zd|\leq 1}D(Z,Z^{\dag})\:
  \exp\left(\bar{\psi}_a^i Z_{ab}\varphi_b^i
    +\bar{\varphi}_a^i\Zd_{ab}\psi_b^i \right)\: 
  {\det}^{|Q|}{\mathcal N}\:, 
\end{align} 
where the $N_c\times N_c$ matrix $\mathcal N$ is defined by ${\mathcal
  N}^{ij}={\bar \varphi}^i_a(1-\Zd Z)_{ab}\varphi^j_b$.  This
completes the derivation of Eq.~\eqref{eq:cft-alg}.

A number of concrete examples illustrating the transformation are
given in App.~\ref{app:ex1}.  In particular, in App.~\ref{app:ex1-div}
we consider an example where the integration measure diverges, and
show how this problem can be solved in a simple case.

\subsection{Generalization to unequal flavor numbers}
\label{sec:unequal}

So far we only considered the case in which particles and
antiparticles have equal flavor numbers, i.e. $N_{b+}=N_{b-}=N_b$,
where $N_{b+}$ ($N_{b-}$) denotes the number of flavors of the
particles (antiparticles).  In practice this constraint may not be
present. It is not difficult to see how our method can be extended to
the general case in which $N_{b+}\neq N_{b-}$. The flavor group is
then $\mathrm U(N_{b+},N_{b-})$, and the integral on the RHS of
Eq.~\eqref{eq:cft-alg} is over the non-compact symmetric space
$\mathrm U(N_{b+}, N_{b-})/[{\mathrm U(N_{b+})\times \mathrm
  U(N_{b-})}]$.  All results derived in earlier parts of this section
are still valid with some minor changes: 1. the complex matrix $Z$ has
dimension $N_{b+}\times N_{b-}$, 2. replace $2N_b$ by $N_{b+}+N_{b-}$
in the integration measure, and 3. choose the range of $Q$
accordingly.  For example, if $N_{b-}<N_c$ and $N_{b+}\ge N_c$, there
are no $Q_-$ sectors, and we sum over $Q\geq 0$ and set $\mathcal N=0$
in our results. We give a concrete example with $N_{b+}\neq N_{b-}$ in
App.~\ref{app:ex2}.

\section{Bosonic color-flavor transformation: character expansion
  method}
\label{sec:char}

In this section, we use the character expansion method~\cite{balant1}
to derive an alternative form of the bosonic color-flavor
transformation which is free from divergences in the range $N_b\leq
N_c$. We will also make use of the results of Refs.~\cite{balant2,ce}.

\subsection{Setup of the calculation}

In the last section, we have traded the integral over the compact
color group for an integral over a non-compact manifold parameterized
by an $N_b\times N_b$ complex matrix $Z$.  Employing a singular value
decomposition, this matrix can be written as
\begin{equation}
  Z=U\Lambda V \:,
\end{equation}
where $U\in {\mathrm U}(N_b)$, $V\in {\mathrm U}(N_b)/{\mathrm
  U^{N_b}(1)}$, and $\Lambda$ is a diagonal matrix with real entries,
the so-called radial coordinates, satisfying $0\le\Lambda_a\le1$.  The
divergence problem we met in the last section is caused by the
integration over the sub-manifold spanned by the radial coordinates.
Specifically, the divergence of highest degree occurs at the boundary,
$\Lambda_a=1$ for all $a$, and the entire information that is needed
to complete the color-flavor transformation resides in the boundary.
A natural question to ask at this point is whether the integration
over the radial coordinates can be avoided.  To answer this question,
we integrate over the two compact unitary groups first.  At the same
time, we relax the constraints on the radial coordinates by replacing
$\Lambda$ with an arbitrary complex matrix.

Our strategy is as follows.  We first perform the integration over the
color group on the LHS of the transformation explicitly
using the character expansion method.  Next we compute an integral
over a compact subgroup of the flavor group with a manifestly
color-invariant integrand.  We then complete the transformation by
observing that the two integrals are equal.

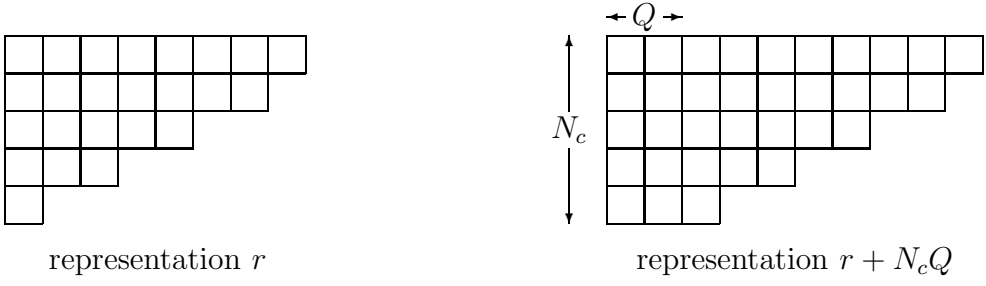
\begin{figure}[tb]
  \begin{center}
    \unitlength1cm
    \begin{picture}(12.5, 4.5)
      \put(8,4.5){\line(1,0){5}}      \put(8,4.0){\line(1,0){5}}
      \put(8,3.5){\line(1,0){4.5}}    \put(8,3.0){\line(1,0){3.5}}
      \put(8,2.5){\line(1,0){2.5}}    \put(8,2.0){\line(1,0){1.5}}
      \put(8.0,2.0){\line(0,1){2.5}}  \put(8.5,2.0){\line(0,1){2.5}}
      \put(9.0,2.0){\line(0,1){2.5}}  \put(9.5,2.0){\line(0,1){2.5}}
      \put(10.0,2.5){\line(0,1){2.0}} \put(10.5,2.5){\line(0,1){2.0}}
      \put(10.5,3.0){\line(0,1){1.5}} \put(11.0,3.0){\line(0,1){1.5}}
      \put(11.5,3.0){\line(0,1){1.5}} \put(12.0,3.5){\line(0,1){1.0}}
      \put(12.5,3.5){\line(0,1){1.0}} \put(13.0,4.0){\line(0,1){0.5}}
      \put(10.5,1.5){\makebox(0,0){representation $r+N_cQ$}}
      \put(7.5,3.5){\vector(0,1){1.0}}
      \put(7.5,3.0){\vector(0,-1){1.0}}
      \put(7.5,3.25){\makebox(0,0){$N_c$}}
      \put(8.25,4.75){\vector(-1,0){0.25}}
      \put(8.75,4.75){\vector(1,0){0.25}}
      \put(8.5,4.75){\makebox(0,0){${Q}$}}

      \put(0,4.5){\line(1,0){4}}     \put(0,4.0){\line(1,0){4}}
      \put(0,3.5){\line(1,0){3.5}}   \put(0,3.0){\line(1,0){2.5}}
      \put(0,2.5){\line(1,0){1.5}}   \put(0,2.0){\line(1,0){0.5}}
      \put(0,2.0){\line(0,1){2.5}}   \put(0.5,2.0){\line(0,1){2.5}}
      \put(1.0,2.5){\line(0,1){2.0}} \put(1.5,2.5){\line(0,1){2.0}}
      \put(2.0,3.0){\line(0,1){1.5}} \put(2.5,3.0){\line(0,1){1.5}}
      \put(3.0,3.5){\line(0,1){1.0}} \put(3.5,3.5){\line(0,1){1.0}}
      \put(4.0,4.0){\line(0,1){0.5}}
      \put(2.1,1.5){\makebox(0,0){representation $r$ }}
    \end{picture}
    \vspace*{-13mm} 
    \caption{Irreducible representations $r$ and $r+N_cQ$.} 
    \vspace*{3mm} 
    \label{fig:young-graph}
  \end{center}
\end{figure}

We define two rectangular $N_c\times N_b$ matrices $\Psi$ and $\Phi$
by
\begin{equation}
  \Psi=(\psi^i_a)\:,\quad \Phi=(\varphi^i_a)\:.  
\end{equation}
The integrand on the LHS of Eq.~\eqref{eq:cft-alg} can then
be rewritten as
\begin{equation}
  \exp\left(\bar{\psi}_{a}^{i}U^{ij}\psi_{a}^{j}+
      \bar{\varphi}_{a}^{i}U^{\dagger ij}\varphi_{a}^{j} \right) =
    \exp\left(\tr UM+ \tr U^{\dagger }N \right)\:, 
\end{equation} 
where we have defined two $N_c\times N_c$ matrices $M$ and $N$ by
\begin{equation}
  \label{eq:mn2}
  M=(M^{ij})=(\psi^i_a\bar {\psi}_{a}^{j})=\Psi\Psi^{\dagger}\:,\quad
  N=(N^{ij})=(\varphi_{a}^{i}\bar \varphi_{a}^{j})=\Phi\Phi^{\dagger}\:.  
\end{equation}
In the following, we consider irreducible representations of $\mathrm
{GL}(m)$ (for various values of $m$) labeled by 
\begin{equation}
  \label{eq:irrep}
  r=(r_1,r_2,\ldots,r_{m})\quad\text{with integers}\quad
  r_1\ge r_2\ge \ldots \ge r_{m}\ge 0\:,
\end{equation}
where $r_j$ is the number of boxes in row $j$ of the corresponding
Young diagram.  
Using Eq.~(3.5) of Ref.~\cite{balant2}, we have
\begin{equation}
  \exp\left( \tr UM\right)=\sum_r \alpha_r^{(0)} \chi_r(UM)\:,\quad
  \exp\left( \tr U^{\dagger}N\right)=\sum_{r'}
  \alpha_{r'}^{(0)}\chi_{r'}(U^{\dagger}N)\:, 
\end{equation}
where the sums are over all irreducible representations of $\mathrm
{GL}(N_c)$ of the form~\eqref{eq:irrep}.  For a given representation
$r$, we have~\cite{balant2}
\begin{align}
\label{eq:alpha}
  \alpha_r^{(\nu)}=\det\left[\frac{1}{(r_j-\nu+i-j)!}\right]
  =\left[\prod^{N_c}_{i=1}\frac{(N_c-i)!}{(k_i-\nu)!}\right]d_{r}
  \quad\text{with}\;\;k_i=N_c+r_i-i\:,
\end{align} 
where $i$ and $j$ run from 1 to $N_c$, $\nu$ is an additional integer
which we shall need later on, $d_r$ is the dimension of the
representation $r$, given by Weyl's formula
\begin{equation}
  \label{eq:dim}  
  d_r=\left[\prod^{N_c-1}_{n=1}n!\right]^{-1}
  \Delta{(k_1,\ldots,k_{N_c})}\:, 
\end{equation}
and $\Delta{(k_1,\ldots,k_{N_c})}=\prod_{i<j}(k_i-k_j)$ is the
Vandermonde determinant.  We then obtain
\begin{align} 
  \label{eq:sum_rr'}
  \int\limits_{{\rm SU}(N_c)}\!\! dU
  \exp\left(\tr UM+ \tr U^{\dagger }N \right)
  &=\sum_{rr'}\alpha_{r}^{(0)}\alpha_{r'}^{(0)}\int\limits_{{\rm
  SU}(N_c)}\!\!dU\ \chi_{r} (UM)\chi_{r'} (U^{\dagger}N)\nn\\
  &=\sum_{rr'}\alpha_{r}^{(0)}\alpha_{r'}^{(0)}\int\limits_{{\rm
  SU}(N_c)}\!\!dU\ U_r^{ij}M_r^{ji}\bar U_{r'}^{kl}N_{r'}^{kl} \:,
\end{align}
where we use the notation $U_r$ for the matrix corresponding to the
representation $r$ specified by a given Young diagram, an example of
which is shown in the left part of Fig.~\ref{fig:young-graph}.
In the right part of that figure, we show a Young diagram that has $Q$
more columns, each containing $N_c$ boxes, than the Young diagram for
$r$.  (Here we assume $Q\geq0$.) We denote the corresponding
representation by $r+N_cQ$.  Note that for $\mathrm {SU}(N_c)$, these
two representations are identical.  We have the orthogonality relation
\begin{align} 
  \label{eq:ortho}
  \int_{{\rm SU}(N_c)}dU\ U_r^{ij}\bar U_{r'}^{kl}
  =\frac{1}{d_r}\delta^{ik}\delta^{jl}\delta_{r',r+N_cQ}\:.
\end{align} 
From Eq.~\eqref{eq:alpha} it is clear that
\begin{equation}
  \alpha_{r+N_cQ}^{(0)}=\alpha_r^{(-Q)}\:.
\end{equation}
Furthermore, for all $g \in \mathrm {GL}(N_c)$ we have \cite{miller}
\begin{align}
  \label{eq:detgQ}
  {g}_{r+N_cQ}^{ij}=g_r^{ij}\:{\det}^Qg\:.
\end{align}
Rewriting the sum over $r$ and $r'$ in Eq.~\eqref{eq:sum_rr'} as a sum
over $r$ and $Q$, we obtain
\begin{align} 
  \label{eq:colorint}
  \int_{{\rm SU}(N_c)}&dU\ \exp\left(\bar{\psi}_{a}^{i}U^{ij}\psi_{a}^{j}+
    \bar{\varphi}_{a}^{i}U^{\dagger ij}\varphi_{a}^{j} \right) \\
  =\sum_r &\:\frac{\alpha_r^{(0)}\alpha_r^{(0)}}{d_r}\chi_{r}(MN)+
  \sum_{Q=1}^{\infty} \left({\det}^Q M+{\det}^Q N
    \right)\sum_{r}\frac{\alpha_{r}^{(0)}\alpha_r^{(-Q)}} 
    {d_r} \chi_{r}(MN)\:.\nn
\end{align} 
The sums over $r$ can be done analytically, resulting in an expression
involving determinants of modified Bessel functions~\cite{ce}, but we
shall not need this explicit result and therefore do not quote it
here.

Note that in the case of color group U($N_c$) only the $Q=0$ term is
non-zero in Eq.~\eqref{eq:ortho}, and therefore only the first term
contributes on the RHS of Eq.~\eqref{eq:colorint}.  This observation
will allow us to read off results for U($N_c$) from those for
SU($N_c$) in Secs.~\ref{sec:nf<nc} and \ref{sec:nf=nc}.

We now turn to the RHS of Eq.~\eqref{eq:cft-alg} and first define two
$N_b\times N_b$ matrices
\begin{equation}
  \label{eq:bc}
    B=(B_{ab})=(\varphi^i_a\bar{\psi}^i_b)=[\Psi^{\dag}\Phi]^T \:, \quad
    C=(C_{ab})=(\psi^i_a\bar{\varphi}^i_b)=[\Phi^{\dag}\Psi]^T \:.
\end{equation}
We then consider the integral
\begin{align} 
  \label{eq:u(f)-int}
  &\int\limits_{\mathrm U(N_b)}\!\!dU \int\limits_{\mathrm U(N_b)}\!\!dV\
  {\det}^{-Q}(UAVB) \exp\left[\bar{\psi}^i_a(UAV)_{ab}\varphi^i_b
    +\bar{\varphi}^i_a(V^{\dagger}DU^{\dagger})_{ab}\psi^i_b \right] \nn\\
  &=\int\limits_{\mathrm U(N_b)}\!\!dU\int\limits_{\mathrm U(N_b) }
  \!\!dV \ {\det}^{-Q}(UAVB)\exp
  \left[\tr(UAVB)+\tr(CV^{\dagger}DU^{\dagger})\right]\:,
\end{align} 
where $A$ and $D$ are two arbitrary $N_b \times N_b$ matrices.  Using
Ref.~\cite{ce}, we have for $AD=I$, i.e. $A$ is the inverse of $D$,
\begin{equation} 
  \label{eq:rhs}
  \eqref{eq:u(f)-int}=
  \sum_{s}\frac{\alpha^{(0)}_s\alpha^{(-Q)}_s}{d^2_s}\chi_s(AD)\chi_s(BC)
  =\sum_{s}\frac{\alpha^{(0)}_s\alpha^{(-Q)}_s}{d_s} \chi_s(BC)\:,
\end{equation}
where the sum is over all irreducible representations $s$ of GL($N_b$)
of the form \eqref{eq:irrep}, and we have used $d_s=\chi_s(I)$.

In App.~\ref{app:proof} we prove the following identity for
$\delta=N_c-N_b\geq0$,
\begin{equation}
  \label{eq:identity}
  \sum_{r}\frac{\alpha_{r}^{(0)}\alpha_r^{(-Q)}}{d_r}
  \chi_{r}(MN)=\prod_{n=0}^{\delta-1}\frac{(N_b+n)!}{(Q+n)!}
  \sum_{s}\frac{\alpha_s^{(0)}\alpha_{s}^{(-Q-\delta)}}{d_{s}}
  \chi_{s}(BC)\:, 
\end{equation}
where the sums on the LHS and the RHS are over all irreducible
representations of GL($N_c$) and GL($N_b$), respectively, that are of
the form~\eqref{eq:irrep}.  Using this identity, we can now relate
Eqs.~\eqref{eq:colorint} and \eqref{eq:rhs}.  We consider separately
the cases $N_b<N_c$, $N_b=N_c$, and $N_b>N_c$.

\subsection{$N_b<N_c$}
\label{sec:nf<nc}

For $N_b<N_c$, the matrices $M$ and $N$ are not of full rank, i.e.\ we
have $\det M=\det N=0$.  The terms multiplied by $\det M$ and $\det N$
in Eq.~\eqref{eq:colorint} are finite~\cite{ce}.  Thus, only the $Q=0$
term in Eq.~\eqref{eq:colorint} survives, and we obtain
\begin{align} 
  \label{eq:nc>nf}
  \eqref{eq:colorint} = \sum_r
  \frac{\alpha_r^{(0)}\alpha_r^{(0)}}{d_r}\chi_{r}(MN) 
  = \prod_{n=0}^{\delta-1}\frac{(N_b+n)!}{n!}\sum_{s}
  \frac{\alpha_s^{(0)}\alpha_{s}^{(-\delta)}}{d_{s}} \chi_{s}(BC)\:,
\end{align}
where we have used Eq.~\eqref{eq:identity}.  This is already in the
form of Eq.~\eqref{eq:rhs} with $Q=\delta=N_c-N_b$.  We can further
simplify the integral in Eq.~\eqref{eq:u(f)-int} by choosing $A=D=I$
and using the invariance of the Haar measure to eliminate $U$ from the
integrand.  This yields
\begin{align}
  &\int_{{\rm SU}(N_c)}dU\ \exp\left(\bar{\psi}_{a}^{i}U^{ij}\psi_{a}^{j}+
    \bar{\varphi}_{a}^{i}U^{\dagger ij}\varphi_{a}^{j} \right) 
  \tag{\ref{eq:nf<nc}}\nn\\
  &=\prod_{n=0}^{N_c-N_b-1}\frac{(N_b+n)!}{n!}
  \int_{\mathrm U(N_b) } dV \ {\det}^{N_b-N_c}
  (VB)\exp\left(\bar{\psi}^i_aV_{ab}\varphi^i_b
      +\bar{\varphi}^i_aV^{\dagger}_{ab}\psi^i_b\right)\nn
\end{align}
as advertised in Sec.~\ref{sec:results}.  If in the above expression
one wants to take the limit of $\det B\to0$, the integral over $V$
needs to be done first.  This procedure yields a finite result,
see the example in App.~\ref{app:e3}.

The result \eqref{eq:nf<nc} is also valid if the integration on the
LHS is over the color group U($N_c$).  This follows immediately from
the remarks made after Eq.~\eqref{eq:colorint} and from the fact that
the terms with $Q>0$ do not contribute on the RHS of
Eq.~\eqref{eq:colorint}.  

\subsection{$N_b=N_c$}
\label{sec:nf=nc}

For $N_b=N_c$ the matrices $M$ and $N$ are of full rank, and all terms
in Eq.~\eqref{eq:colorint} contribute.  Eq.~\eqref{eq:identity} now
becomes trivial,
\begin{equation}
  \sum_{r}\frac{\alpha_{r}^{(0)}\alpha_r^{(-Q)}}{d_r} \chi_{r}(MN)=
  \sum_{s}\frac{\alpha_s^{(0)}\alpha_{s}^{(-Q)}}{d_{s}} \chi_{s}(BC)\:.
\end{equation}
We again simplify the integral in Eq.~\eqref{eq:u(f)-int} by choosing
$A=D=I$ and using the invariance of the Haar measure to arrive at
\begin{align} 
  \int_{\mathrm {SU}(N_c)} d&U\:\exp\left(\bar{\psi}_{a}^{i}U^{ij}\psi_{a}^{j}
    +\bar{\varphi}_{a}^{i}U^{\dagger ij}\varphi_{a}^{j}\right) \nn \\
  &=\sum_{Q=0}^{\infty}\tilde \chi_Q \int_{\mathrm U(N_b)} dV \ 
  {\det}^{-Q}(VB)\exp \left(\bar{\psi}^i_aV_{ab}\varphi^i_b+
    \bar{\varphi}^i_aV^{\dagger}_{ab}\psi^i_b\right)
  \tag{\ref{eq:nf=nc}}
\end{align}
with $\tilde \chi_0=1$ and $\tilde \chi_{Q>0}={\det}^QM+{\det}^QN$ as
stated in Sec.~\ref{sec:results}.  If the $\det B\to0$ limit is
desired, the integral over $V$ needs to be done first as mentioned at
the end of the previous subsection.  If ${\det}^{-Q}(VB)$ is combined
with the terms in $\tilde \chi_{Q>0}$, we obtain
$\beta^Q+1/\bar\beta^Q$ with $\beta=\det\Psi/\det(\Phi V)$.  This
shows that we are not allowed to change the order of summation and
integration in Eq.~\eqref{eq:nf=nc}, since the resulting geometric
series would diverge for one of the two terms.

If the integration on the LHS of Eq.~\eqref{eq:nf=nc} is over the
color group U($N_c$), only the $Q=0$ term contributes on the RHS as
explained after Eq.~\eqref{eq:colorint}, and we obtain for $N_b=N_c$
\begin{equation}
  \label{eq:Uint}
  \int_{\mathrm {U}(N_c)} dU\:
  \exp\left(\bar{\psi}_{a}^{i} U^{ij}\psi_{a}^{j}
    +\bar{\varphi}_{a}^{i}U^{\dagger ij}\varphi_{a}^{j}\right)
  =\int_{\mathrm U(N_b)} dV \: 
  \exp \left(\bar{\psi}^i_aV_{ab}\varphi^i_b
    + \bar{\varphi}^i_aV^{\dagger}_{ab}\psi^i_b\right)\:,
\end{equation}
see also Sec.~6 of Ref.~\cite{induce}.

\subsection{$N_b>N_c$}

In this case the $N_c\times N_c$ matrices $M$ and $N$ are of full
rank, whereas the $N_b\times N_b$ matrices $B$ and $C$ are of rank
$N_c$ with $N_b-N_c$ eigenvalues equal to zero so that $\det B=\det
C=0$.  We now have $\delta=N_c-N_b<0$.  Using similar arguments as in
App.~\ref{app:proof} but in the reverse direction, we obtain instead
of Eq.~\eqref{eq:identity}
\begin{align} 
  \label{eq:nf>nc}  
  & \sum_{r}\frac{\alpha_{r}^{(0)}\alpha_{r}^{(-Q)}}{d_r} \chi_{r}(MN)
  =C_{|\delta|}\sum_{s}\frac{\alpha_{s}^{(0)}\alpha_{s}^{(|\delta|-Q)}}
  {d_{s}}\chi_{s}(BC)\nn\\ 
  &=C_{|\delta|}\!\int_{\mathrm U(N_b)} \!dU\int_{\mathrm
    U(N_b) } \!dV \ {\det}^{|\delta|-Q}(UAVB)
  \exp\left[\tr(UAVB)+\tr(CV^{\dagger}DU^{\dagger}) \right]\nn\\
  &=C_{|\delta|}\!\int_{\mathrm U(N_b) } dV \
  {\det}^{|\delta|-Q}(VB)\exp\left[\tr(VB)+\tr(CV^{\dagger})\right]
\end{align} 
with 
\begin{equation}
  C_{|\delta|}=\prod_{n=1}^{|\delta|}\frac{(Q-n)!}{(N_b-n)!}\:,
\end{equation}
where in the last step in Eq.~\eqref{eq:nf>nc} we have again set
$A=D=I$ and used the invariance of the Haar measure to eliminate $U$
from the integrand.  Note that this expression is only valid for
$Q\geq|\delta|=N_b-N_c$.  Although $\det B=0$ appears with a
non-positive power, the RHS of Eq.~\eqref{eq:nf>nc} must
be finite because the LHS is.  This fact can be established
explicitly by a suitable limiting procedure.

For $0\leq Q<|\delta|$, the integral over U($N_b$) in
Eq.~\eqref{eq:nf>nc} is zero because $\det B=0$ appears with a
positive power.  For this range of $Q$, we cannot replace the
corresponding terms in Eq.~\eqref{eq:colorint} by integrals over
U($N_b$) and therefore cannot complete the transformation.  Thus, it
seems that the character expansion method does not yield a useful
result for $N_b>N_c$.

\section{Conclusions and outlook}
\label{sec:concl}

We have generalized Zirnbauer's color-flavor transformation in the
bosonic sector to the special unitary group $\mathrm {SU}(N_c)$.
Because the flavor group U($N_b,N_b$) is non-compact, divergences
arise if the number of bosonic flavors is too large.  This has already
been noted in Refs.~\cite{Zirnbauer:1997qm,induce} where the gauge
group was U($N_c$) and results were given for $2N_b\le N_c$.  We have
found a ``standard'' result for the color-flavor transformation in the
same range, and an alternative form of the transformation in the
extended range $N_b\le N_c$.  (A special case of this result for
$N_b=N_c$ and color group U($N_c$) has already been given in
Ref.~\cite{induce}.)  For $N_b<N_c$, the results for SU($N_c$) are
identical to those for U($N_c$) because only the sector with $Q=0$
contributes.

The results of the present paper can be applied to study a
boson-induced SU($N_c$) lattice gauge theory analogous to the U($N_c$)
gauge theory discussed in Ref.~\cite{induce}.  We hope that other
applications will arise, e.g., in the field of disordered and/or
chaotic systems.

One obvious open problem is to obtain a manifestly convergent result
for $N_b>N_c$.  While the divergences that appear in numerator and
denominator of our formal result \eqref{eq:cft-alg} can always be
canceled in special cases, see App.~\ref{app:ex1-div}, the general
case is difficult to deal with.  The character expansion method, which
led to a convergent result in the extended range $N_b\le N_c$, fails
for $N_b>N_c$ since it cannot generate the terms with $0\le Q<N_b-N_c$
on the RHS of Eq.~\eqref{eq:colorint} in terms of integrals over (a
subgroup of) the flavor group.  However, as stated earlier and in
Ref.~\cite{induce}, all necessary information resides in the boundary
of the coset space U($N_b,N_b$)/[U($N_b)\times$U($N_b$)], so it is
conceivable that an explicit result in terms of an integration over
this boundary might yet be obtained.

The other open problem is the extension of the present results to the
supersymmetric case in which both fermionic and bosonic flavors are
present.  In this case the divergence of the integration measure due
to the bosonic degrees of freedom can be canceled by the contribution
of the fermions to the measure, as long as sufficiently many fermions
are included.  The physically interesting case is $N_c=3$ (the gauge
group of QCD), $N_f\ge2$ (the number of physical quark flavors), and
$N_b=N_c$ (the lower bound for $N_b$ so that the bosons induce the
correct gauge action \cite{induce}).  In this case, the convergence
requirement $2(N_b-N_f)\le N_c$ is satisfied.  However, the
supersymmetric case raises other issues which will be addressed in a
separate publication.

\section*{Acknowledgements}
This work was supported in part by the U.S.\ Department of Energy
(contract no.\ DE-FG02-91ER40608), by the RIKEN-BNL Research Center,
and by Deutsche Forschungsgemeinschaft (project no. FOR 465).  We
would like to thank B.\ Schlittgen for many helpful conversations and
M.R.\ Zirnbauer for a stimulating discussion.

\appendix

\section{Normalization of generalized Slater states}
\label{app:norm}

In this section, we calculate the norm, $N_Q=\langle \psi_Q|\psi_Q
\rangle$, of the state $|\psi_Q\rangle$ defined in
Eq.~\eqref{eq:psiq}.  We first assume $Q\ge0$ and discuss the case of
$Q<0$ at the end of this section.  The vacuum is normalized by
definition so that $N_0=1$.  For $Q=1$, $|\psi_Q\rangle$ is the Slater
state with the well-known norm $N_1=N_c!$.

To prove Eq.~\eqref{eq:nq}, we study a different version of the
color-flavor transformation in which the flavor group is ${\mathrm
  U}(N_b)$ with $N_b\!=\!N_c$.  Note that the flavor group is compact
now.  We follow the same method as in Sec.~\ref{sec:alg} but keep only
the particles created by the $\cba^i_a$ and discard the antiparticles
created by the $\dba^i_a$.  The flavor group ${\mathrm U}(N_b)$ is
then generated by $\{\tilde {\mathcal G}_{ab}=\cba^i_ac^i_b\}$. The
state $|\psi_Q \rangle$ is still defined as in Eq.~\eqref{eq:psiq}.
The projector onto the $Q$-sector is now
\begin{align}
  \1_Q =\tilde C_Q\int_{{\mathrm U}(N_b)} dg \ \tilde
  T_g|\psi_{Q}\rangle \langle \psi_{Q}|\tilde T_{g}^{-1}
\end{align}
with the normalized Haar measure $dg$ of U($N_b$) and $\tilde
T_g=\exp({\cba}^i_a(\ln g)_{ab}c^i_b)$.  For $N_b=N_c$, we have by
explicit calculation
\begin{align}
  \label{eq:detg}
  \tilde T_g|\psi_Q\rangle=({\det}^Qg)\:|\psi_Q\rangle\:.
\end{align}
Using this equation, the normalization constant $\tilde C_Q$, see also
Eq.~\eqref{eq:cqm}, simplifies to
\begin{align}
  \tilde C_Q=N_Q\left[\int_{{\mathrm U}(N_b)} dg\ 
  \langle \psi_Q| \tilde T_g|\psi_{Q}\rangle \langle\psi_{Q}|\tilde
  T_{g}^{-1}|\psi_Q\rangle\right]^{-1} 
  \!\!=N_Q\left[N_Q^2\right]^{-1}\!=\frac1{N_Q}\:.
\end{align}
Next we consider the following integral and perform manipulations
similar to those in Sec.~\ref{sec:alg},
\begin{align} 
  \label{eq:int1}
  \int_{\mathrm {SU}(N_c)} dU &
  \exp\left(\bar{\psi}_{a}^{i}U^{ij}\psi_{a}^{j} \right)
  =\int_{\mathrm {SU}(N_c)} dU\  \langle0|\exp(\bar\psi^i_ac^i_a)
  \exp\left(\bar{c}_{a}^{i}U^{ij}\psi_{a}^{j}\right) |0\rangle \nn\\
  &=\sum_{Q=0}^\infty \ \langle 0|\exp(\bar\psi^i_ac^i_a ) \1_Q \exp
  (\cba^i_a \psi ^i_a )|0\rangle \nn\\
  &=\sum_{Q=0}^\infty \tilde C_Q \int_{{\mathrm U}(N_b)} dg \ 
  \langle 0|\exp(\bar\psi^i_ac^i_a )\tilde T_g|\psi_{Q}\rangle 
  \langle\psi_{Q}|\tilde T_{g}^{-1}\exp (\cba^i_a \psi ^i_a )|0\rangle \nn\\
  &=\sum_{Q=0}^\infty \frac{1}{N_Q}\langle 0|\exp(\bar\psi^i_ac^i_a
  )|\psi_{Q}\rangle \langle\psi_{Q}|\exp (\cba^i_a \psi ^i_a )|0\rangle \nn\\
  &=\sum_{Q=0}^\infty \frac{1}{N_Q}\: {\det}^Q M \:,
\end{align} 
where $M$ is an $N_c\times N_c$ matrix, $M^{ij}=\psi^i_a
\bar\psi^j_a$, and we have again used Eq.~\eqref{eq:detg}.  However,
we can also use the character expansion method~\cite{balant1,balant2}
to do this integral.  Using the same notation as in
Sec.~\ref{sec:char}, we obtain
\begin{align} 
  \label{eq:int2}
  \int_{\mathrm {SU}(N_c)} dU
  \exp\left(\bar{\psi}_{a}^{i}U^{ij}\psi_{a}^{j} \right) 
  &=\int_{\mathrm {SU}(N_c)} dU \exp\left(\tr UM \right)\nn \\
  &=\int_{\mathrm {SU}(N_c)} dU \sum_r \alpha_r^{(0)} \chi_r (UM)\nn \\
  &= \sum_r \ \alpha_r^{(0)} M_r^{ji} \int_{\mathrm {SU}(N_c)} dU \
  U_r^{ij}\nn\\ 
  &= \sum_{Q=0}^\infty \alpha_{r=N_cQ}^{(0)}\: {\det}^Q M\:.
\end{align} 
In the last step, we have used the facts that 
\begin{align} 
  \label{eq:Uab}
  \int_{\mathrm {SU}(N_c)}dU\  U_r^{ij} =\left\{ 
    \begin{array}{ll}
      1\:, &\quad r=N_cQ\:, \\
      0\:, &\quad \text{else}\:,
    \end{array}\right.
\end{align}
where $r=N_cQ$ denotes the (one-dimensional) irreducible
representation of GL($N_c$) specified by a Young diagram with $N_c$
rows and $Q$ columns, and that for all $M \in \mathrm{GL}(N_c)$ we
have $M_{r=N_cQ}^{ij}={\det}^QM$, see Eq.~\eqref{eq:detgQ}.  (Note
that in the one-dimensional representation $r=N_cQ$ the indices $i$
and $j$ only take the value 1.)  From Eq.~\eqref{eq:alpha} we obtain
with $d_{r=N_cQ}=1$
\begin{equation}
  \alpha_{r=N_c Q}^{(0)}=\prod^{N_c-1}_{n=0}\frac{n!}{(Q+n)!}\:.
\end{equation}
Comparing Eqs.~\eqref{eq:int1} and \eqref{eq:int2} we arrive at
Eq.~\eqref{eq:nq}, valid for $Q\ge0$.  The calculation for $Q<0$
proceeds in complete analogy by working with the antiparticles instead
of the particles, and the result for this case can be obtained 
by the replacement $Q\to-Q$ in the expression for $N_Q$.

\section{Calculation of the $\boldsymbol{\mathcal C_Q}$}
\label{app:rad}

In this section, we do the integral in Eq.~\eqref{eq:radial-int}.
This is an example of so-called Hua-type integrals that were studied
by Hua a long time ago~\cite{hua} and recently extended by
Neretin~\cite{neretin}.  Here, we follow the method introduced in
Ref.~\cite{hua}.  We first consider the case of $N_b\ge N_c$ and
$Q\ge0$, and then give a result for $N_b<N_c$ and $Q=0$.

Using Eq.~\eqref{eq:Dz}, Eq.~\eqref{eq:radial-int} becomes
\begin{align} 
  {\mathcal C}_Q^{-1}={N_Q} \int_{|Z\Zd|\leq 1}\frac{dZdZ^{\dagger}}
  {{\det}^{2N_b-N_c}(1-ZZ^{\dagger})}{\det}^Q(1-Z\Zd)_{[N_c]}\:.
\end{align}
We now write the matrix $Z$ as $Z=(Z_{N_b,N_b-1},q)$, where
$Z_{N_b,N_b-1}$ is an $N_b\times (N_b-1)$ matrix and $q$ is a single
column. We then have
\begin{equation}
  1-ZZ^\dagger=1-Z_{N_b,N_b-1}Z_{N_b,N_b-1}^{\dagger}-qq^\dagger\:.
\end{equation}
Since $|1-Z_{N_b,N_b-1}Z_{N_b,N_b-1}^{\dagger}|\geq 0$, i.e.\ the
matrix has real and non-negative eigenvalues, we write
$1-Z_{N_b,N_b-1}Z_{N_b,N_b-1}^{\dagger}=\Gamma\Gamma^{\dagger}$ and
define $q=\Gamma w$. Then
\begin{equation}
  dqdq^\dagger=|\det \Gamma |^2dwdw^\dagger=\det
  \left(1-Z_{N_b,N_b-1}Z_{N_b,N_b-1}^{\dagger}\right)dwdw^\dagger \:.
\end{equation}
On the other hand,
\begin{align*}
  \det(1-ZZ^{\dagger})=\det\left[\Gamma(1-ww^\dagger)\Gamma^{\dagger}\right]
    =(1-w^\dagger w)\det(1-Z_{N_b,N_b-1}Z_{N_b,N_b-1}^{\dagger}) \:,
\end{align*} 
where we have used $\det(1-ww^\dagger)=(1-w^\dagger w)$.  Applying the
same procedure to $Z_{N_b,N_b-1}$ and so on, we obtain
\begin{align} 
  &[\mathcal C_Q N_Q]^{-1}=\int_{|Z\Zd|\leq 1}\frac{ dZdZ^{\dagger}}
  {{\det}^{2N_b-N_c}(1-ZZ^{\dagger})}{\det}^Q(1-Z\Zd)_{[N_c]} \nn \\
  &=\prod_{i=1}^{N_c}\int\limits_{\;w_i^\dagger w_i\leq 1} \hspace{-3mm} dw_i 
  dw_i^\dagger(1-w_i^\dagger w_i)^{N_c-N_b+Q-i} \!\!\! \prod_{j=N_c+1}^{N_b}
  \int\limits_{\;w_j^\dagger w_j\leq 1} \hspace{-3mm} dw_jdw_j^\dagger
  (1-w_j^\dagger w_j)^{N_c-N_b-j}\nn\\
  &=\prod_{i=1}^{N_c}\pi^{N_b}\frac{(N_c-N_b+Q-i)!}{(N_c+Q-i)!}
  \prod_{j=N_c+1}^{N_b}\pi^{N_b}\frac{(N_c-N_b-j)!}{(N_c-j)!}\:.
\end{align}
From this equation, which was derived for $N_b\ge N_c$, we see that
for $N_b>N_c$, the above integral diverges for all $Q\ge0$, whereas
for $N_b=N_c$, it diverges for $Q<N_b$.

For $N_b<N_c$, only the $Q=0$ sector exists, and we obtain from a very
similar and even simpler calculation
\begin{equation}
  \label{eq:C0}
  {\mathcal C}_0^{-1}=C_0^{-1}=\pi^{N_b^2}\prod_{n=1}^{N_b}
  \frac{(N_c-N_b-n)!}{(N_c-n)!}\:,
\end{equation}
which is finite for $2N_b\le N_c$ but diverges for $N_b<N_c<2N_b$.

\section{Proof of identity (\ref{eq:identity})}
\label{app:proof}

We have $\tr (MN)=\tr (\Psi\Psi^\dagger\Phi\Phi^\dagger)=\tr
(\Phi^\dagger\Psi\Psi^\dagger\Phi)=\tr
(\Phi^\dagger\Psi\Psi^\dagger\Phi)^T=\tr (BC)$, which proves the
identity immediately for $N_b=N_c$.  In the following, we assume
$\delta=N_c-N_b>0$ and prove the identity by an iterative procedure.

The semi-positive definite $N_b\times N_b$ matrix
$BC=(\Phi^\dagger\Psi\Psi^\dagger\Phi)^T$ has $N_b$ eigenvalues that
we denote by $\lambda_a^2$ with $a=1,\ldots,N_b$.  From linear
algebra~\cite{lc} we know that the $N_c\times N_c$ matrix
$MN=\Psi\Psi^\dagger\Phi\Phi^\dagger$ has $N_b$ eigenvalues equal to
those of $BC$ and $\delta=N_c-N_b$ eigenvalues equal to zero.  We
denote the eigenvalues of $MN$ by $\mu_i^2$, with $\mu_i=\lambda_i$
for $1\leq i\leq N_b$.  For the purpose of this proof, we start with
nonzero values $\mu_i$ for $N_b<i\leq N_c$ and let them go to
zero one by one, starting with $\mu_{N_c}$. Weyl's character formula
is \cite{hua}
\begin{equation}
  \chi_{(r_1, r_2, \ldots, r_{N_c})}(MN)=\frac{\det(\mu^{2(N_c+r_j-j)}_{i})}
  {\Delta(\mu^2)}=\frac{\det(\mu_i^{2k_j})}{\Delta(\mu^2)}\:,
\end{equation}
where the $k_j$ have been defined in Eq.~\eqref{eq:alpha} and $\Delta$
again denotes the Vandermonde determinant.  We have
\begin{equation}
  \label{eq:delta}
  \lim_{\mu_{N_c}\to 0}\Delta_{N_c}(\mu^2)=
  \lim_{\mu_{N_c}\to 0}\prod_{i<j}^{N_c}
  (\mu_i^2-\mu_j^2)=\Delta_{N_c-1}(\mu^2) \prod^{N_c-1}_{i=1}\mu^2_i\:, 
\end{equation}
where the index on $\Delta$ denotes the number of eigenvalues
involved.  Similarly, we introduce the notation
$\det_m(\mu_i^{2k_j})$ to denote the determinant of the $m\times m$
upper-left sub-matrix of the matrix $(\mu_i^{2k_j})$ , i.e.\ $i$ and
$j$ run from 1 to $m$ instead of from 1 to $N_c$.  Note that for
$\mu_{N_c}\to0$, $\det_{N_c}(\mu^{2k_j}_i)$ is non-vanishing only if
$k_{N_c}=0$ or, equivalently, $r_{N_c}=0$, in which case we have
\begin{align}
  \label{eq:det} 
  \lim_{\mu_{N_c}\to 0}{\det}_{N_c} (\mu^{2k_j}_i)=
    {\det}_{N_c-1}(\mu^{2k'_j}_i) \prod^{N_c-1}_{i=1} \mu^2_i\:,
\end{align} 
where
\begin{equation}
  k'_j=r_j+(N_c-1)-j \quad\text{with}\quad j=1, \ldots, N_c-1\:.
\end{equation}
We conclude that for $\mu_{N_c}=0$, $\chi_r(MN)$ is non-zero only for
representations of GL($N_c$) with Young diagram $r=(r_1\geq \cdots\geq
r_{N_c}=0)$.  We thus obtain
\begin{align} 
  \lim_{\mu_{N_c}\to 0}&\sum_{r}\frac{\alpha_{r}^{(0)}
    \alpha_r^{(-Q)}}{d_r} \chi_{r}(MN)\nn\\ 
  &=\sum_{r_1\geq \ldots \geq r_{N_c=0}}
  \frac{\alpha_{r}^{(0)}\alpha_{r}^{(-Q)}}{d_r}\frac{{\det}_{N_c-1}
  (\mu^{2k'_j}_i)}{\Delta_{N_c-1}(\mu^2)}\\
  &=\prod\limits_{n=1}^{N_c-1}n! \sum_{k_1>\ldots>k_{N_c}=0}
  {\det}_{N_c}\Bigl(\frac{1}{k_j!(k_j-N_c+Q+i)!}\Bigr)
  \frac{{\det}_{N_c-1}(\mu^{2k'_j}_i)}{\Delta_{N_c-1}(\mu^2)}\:, \nn
\end{align}
where we have used Eqs.~\eqref{eq:alpha} and \eqref{eq:dim}.
From Eq.~\eqref{eq:alpha} with $k_{N_c}=0$ we have 
\begin{align}
  \label{eq:detn}
  {\det}_{N_c}\Bigl(\frac{1}{k_j!(k_j-N_c+Q+i)!}\Bigr)
  = \frac{d^{\text{GL}(N_c)}_{(r_1,\ldots,r_{N_c-1},0)}}
  {\prod\limits^{N_c-1}_{j=1}k_j!}\prod_{i=1}^{N_c}\frac{(N_c-i)!}
  {(k_i+Q)!}\:.
\end{align} 
From Eq.~\eqref{eq:dim} we find that
\begin{equation}
  \label{eq:dndn-1}
  d^{\text{GL}(N_c)}_{(r_1,\ldots,r_{N_c-1},0)}
  =\frac{\prod\limits^{N_c-1}_{j=1}k_j}{(N_c-1)!}\
  d^{\text{GL}(N_c-1)}_{(r_1+1, \ldots, r_{N_c-1}+1)}\:.
\end{equation}
Hence, for $\mu_{N_c}=0$, and thus $k_{N_c}=0$, we obtain from
\eqref{eq:detn}, \eqref{eq:dndn-1}, and \eqref{eq:alpha}
\begin{align}
  {\det}_{N_c}\Bigl(&\frac{1}{k_j!(k_j-N_c+Q+i)!}\Bigr)\nn\\
  &=\frac{1}{Q!}\:{\det}_{N_c-1}\Bigl(\frac{1}{k'_j!
    (k'_j-(N_c-1)+(Q+1)+i)!}\Bigr)\:.
\end{align} 
Putting everything together, we arrive at
\begin{equation}
 \lim_{\mu_{N_c}\to
 0}\sum_{r}\frac{\alpha_{r}^{(0)}\alpha_{r}^{(-Q)}}{d_r} \chi_{r}(MN)= 
 \frac{(N_c-1)!}{Q!}\sum_{r'}\frac{\alpha_{r'}^{(0)}\alpha_{r'}^{(-Q-1)}}
 {d_{r'}} \chi_{r'}(MN)\:,
\end{equation}
where the sum on the RHS is over all irreducible representations
$r'=(r_1\geq\ldots\geq r_{N_c-1}\geq 0)$ of GL($N_c-1$).  Repeating
this procedure $\delta=N_c-N_b$ times, we obtain our identity
\begin{equation}
  \tag{\ref{eq:identity}}
  \sum_{r}\frac{\alpha_{r}^{(0)}\alpha_{r}^{(-Q)}}{d_r}
  \chi_{r}(MN)=\prod_{n=0}^{\delta-1}\frac{(N_b+n)!}{(Q+n)!}
  \sum_{s}\frac{\alpha_{s}^{(0)}\alpha_{s}^{(-Q-\delta)}}{d_{s}}
  \chi_{s}(BC)\:, 
\end{equation}
where the sum on the RHS is over all irreducible representations $s$
of GL($N_b$) of the form~\eqref{eq:irrep}, and where we have again
used $\tr(MN)=\tr(BC)$.

\section{Examples for the algebraic result}

\subsection{$N_c=2$, $N_b=1$}
\label{app:ex1}

We parameterize elements of SU(2) as
\begin{equation}
  \label{eq:SU2}
  U=\begin{pmatrix}
    e^{i\lambda}\cos\theta & -e^{i\eta}\sin\theta \\
    e^{-i\eta}\sin\theta &e^{-i\lambda}\cos\theta
  \end{pmatrix} \quad\text{with }
  0\leq\theta\leq\frac{\pi}{2}\:,\; 0\leq \lambda,\eta<2\pi\:.
\end{equation}
The corresponding normalized Haar measure is
$dU\!=\!(1/2\pi^2)\sin\theta\cos\theta d\theta d\lambda d\eta$.
Performing the integral on the LHS of
Eq.~\eqref{eq:cft-simple}, we obtain
\begin{align}
  \label{eq:lhs}
  &\int_{{\rm SU}(2)} dU \exp\left(\bar{\psi}^{i}U^{ij}\psi^{j}+
    \bar{\varphi}^{i}U^{\dagger ij}\varphi^{j} \right) \nn \\
  &= \sum_{n=0}^{\infty}\frac{1}{n!(n+1)!}(
    {\bar \psi}^1\psi^1{\bar \varphi}^1{\varphi}^1+
    {\bar \psi}^2\psi^2{\bar \varphi}^2{\varphi}^2+
    {\bar \psi}^1\psi^2{\bar \varphi}^2{\varphi}^1+
    {\bar \psi}^2\psi^1{\bar \varphi}^1{\varphi}^2)^n\:.
\end{align}
For the RHS of Eq.~\eqref{eq:cft-simple}, we have with
$\mathcal C_0=1/\pi$
\begin{align}
  &\frac{1}{\pi}\int_{|z|\leq1}dzd\bar z\exp\left[
    ({\bar \psi}^1\varphi^1+{\bar \psi}^2\varphi^2)z+
    ({\bar \varphi}^1\psi^1+{\bar \varphi}^2\psi^2)\bar z\right] \nn \\
  &=\frac{1}{\pi}\int^1_0 rdr\int^{2\pi}_0d\theta\exp\left[
    ({\bar \psi}^1\varphi^1+{\bar \psi}^2\varphi^2)re^{i\theta}+
    ({\bar \varphi}^1\psi^1+{\bar \varphi}^2\psi^2)re^{-i\theta}
  \right]\nn \\
  &=\sum_{n=0}^{\infty}\frac{1}{n!(n+1)!}({\bar \psi}^1\psi^1{\bar
    \varphi}^1{\varphi}^1+ {\bar \psi}^2\psi^2{\bar
    \varphi}^2{\varphi}^2+{\bar \psi}^1\psi^2{\bar
    \varphi}^2{\varphi}^1+{\bar \psi}^2\psi^1{\bar
    \varphi}^1{\varphi}^2)^n
\end{align}
in agreement with Eq.~\eqref{eq:lhs}.

\subsection{$N_c=N_{b+}=2$, $N_{b-}=0$}
\label{app:ex2}

In this example we check the argument we made in
Sec.~\ref{sec:unequal} for $N_{b+}\neq N_{b-}$.  To have $N_{b-}=0$ we
simply set ${\bar \varphi}^i_a=\varphi^i_a=0$.  Using the same
parameterization for SU(2) as in App.~\ref{app:ex1}, we calculate the
integral on the LHS of Eq.~\eqref{eq:cft-alg},
\begin{align}
  \label{eq:lhs2}
  &\int_{\rm {SU}(2)} dU
  \exp\left(\bar{\psi}_{a}^{i}U^{ij}\psi_{a}^{j} \right) \nn \\
  &=\sum_{n=0}^{\infty}\frac{1}{n!(n+1)!}({\bar
    \psi}^1_1\psi^1_1{\bar \psi}^2_2\psi^2_2+{\bar
    \psi}^2_1\psi^2_1{\bar \psi}^1_2\psi^1_2-{\bar
    \psi}^1_1\psi^2_1{\bar \psi}^2_2\psi^1_2-{\psi}^1_1\bar
  \psi^2_1{\bar \psi}^1_2\psi^2_2)^n\:.
\end{align}
In this case we have to sum over $Q$ from zero to infinity on the RHS.
The integral over the coset space U(2)/U(2) amounts to evaluating the
integrand at the single point $Z=0$.  From Eqs.~\eqref{eq:radial-int}
and \eqref{eq:nq}, we have $\mathcal C_Q=1/N_Q=1/[Q!(Q+1)!]$ and thus
obtain
\begin{align}
  \text{RHS}&=\sum_{Q=0}^{\infty}\frac{1}{Q!(Q+1)!}
  {\det}^Q{\mathcal M}\\
  &= \sum_{Q=0}^{\infty}\frac{1}{Q!(Q+1)!}({\bar
    \psi}^1_1\psi^1_1{\bar \psi}^2_2\psi^2_2+{\bar
    \psi}^2_1\psi^2_1{\bar \psi}^1_2\psi^1_2-{\bar
    \psi}^1_1\psi^2_1{\bar \psi}^2_2\psi^1_2-{\psi}^1_1\bar
  \psi^2_1{\bar \psi}^1_2\psi^2_2)^Q \nn
\end{align}
in agreement with Eq.~\eqref{eq:lhs2}, where we have used ${\mathcal
  M}^{ij}={\bar \psi}^i_a\psi^j_a$.

\subsection{$N_c=N_b=1$}
\label{app:ex1-div}

In this example, we run into the divergence problem discussed in
Secs.~\ref{sec:results} and \ref{sec:alg}.  The LHS of
Eq.~\eqref{eq:cft-alg} is simple because the integral over $\mathrm
{SU}(1)$ reduces to evaluating the integrand at unity,
\begin{equation}
  \int_{{\rm SU}(1)} dU \exp\left(\bar{\psi}^{i}U^{ij}\psi^{j}+
    \bar{\varphi}^{i}U^{\dagger ij}\varphi^{j} \right)
  =\exp(\bar\psi\psi+\bar\varphi\varphi)\:.
\end{equation}
The RHS of Eq.~\eqref{eq:cft-alg} is a sum over $Q$. For
$Q=0$ we have
\begin{equation}
  {\mathcal C}_0^{-1}
  =\int_{|z|\leq 1}\frac{dzd\bar z}{1-z\bar z}
  =2\pi\int_{0}^1\frac{rdr}{1-r^2}
\end{equation}
and
\begin{equation}
  \int_{|z|\leq 1}\frac{dzd\bar z}{1-z\bar z}
  \exp(\bar\psi z\varphi+\bar\varphi \bar z\psi)
  =2\pi\sum_{n=0}^\infty\frac{(\bar\psi\psi\bar\varphi\varphi)^n}{(n!)^2}
  \int_{0}^1\frac{r^{2n+1}dr}{1-r^2} \:.
\end{equation}
We now change the upper limit in the integral to $1-\epsilon$ and
let $\epsilon\to0$ to obtain
\begin{equation}
  {\mathcal C}_0\int_{|z|\leq 1}\frac{dzd\bar z}{1-z\bar z}
  \exp(\bar\psi z\varphi+\bar\varphi \bar z\psi)
  =\sum_{n=0}^\infty\frac{(\bar\psi\psi\bar\varphi\varphi)^n}{(n!)^2} \:.
\end{equation}
For $Q\geq1$, there are no divergences, and we have
\begin{align} 
  \int_{|z|\leq 1}\frac{dzd\bar z}{(1-z\bar z)^{2N_b-N_c-Q}}
  \exp(\bar\psi z \varphi+\bar\varphi \bar z\psi)
  =\pi\sum_{n=0}^\infty\frac{(Q-1)!}{n!(n+Q)!}
  (\bar\psi\psi\bar\varphi\varphi)^n\:.
\end{align} 
Collecting all terms and using ${\mathcal C}_Q=1/[\pi(Q-1)!]$ , we
find for the RHS of Eq.~\eqref{eq:cft-alg}
\begin{align}
  \label{eq:sumnQ}
  &\int_{|ZZ^{\dagger}|\leq 1} D(Z,\Zd) 
  \exp\left(\bar{\psi}_a^i Z_{ab}\varphi_b^i
    +\bar{\varphi}_a^i\Zd_{ab}\psi_b^i \right) \sum_{Q=0}^{\infty}
  \chi_Q\\
  &=\sum_{n=0}^\infty\sum_{Q=0}^\infty\frac{1}{n!(n+Q)!}
  \left[(\bar\psi\psi)^Q+(\bar\varphi\varphi)^Q-\delta_{Q0}\right]
  (\bar\psi\psi\bar\varphi\varphi)^n 
  =\exp(\bar\psi\psi+\bar\varphi\varphi)\:,\nn
\end{align}
where the last step requires some rearrangements of the terms in the
sums.  We see that the transformation works although for $Q=0$ the
normalization factor and the integral on the RHS of the transformation
are divergent.  After those infinities have been canceled, the finite
ratio gives the correct result.  However, as mentioned at the end of
Sec.~\ref{sec:Z0}, it is not trivial to obtain a simple result for the
general case.

\section{Examples for the character expansion result}

\subsection{$N_c=N_b=1$}

Although this example was already considered in
App.~\ref{app:ex1-div}, we revisit it here to check our result
obtained using the character expansion method.  Again, the LHS equals
$\exp(\bar\psi\psi+\bar\varphi\varphi)$. From Eq.~\eqref{eq:nf=nc} we
have
\begin{align} 
  \sum_{Q=0}^\infty&\tilde\chi_Q\int_0^{2\pi}\frac{d\theta}{2\pi}
  (Be^{i\theta})^{-Q}\ \exp\left(\bar\psi e^{i\theta}\varphi
    +\bar\varphi e^{-i\theta}\psi\right)\\
  &=\sum_{Q=0}^\infty\left[(\bar\psi\psi)^Q+(\bar\varphi\varphi)^Q
  -\delta_{Q0}\right]\sum_{n=0}^\infty\frac{1}{n!(n+Q)!}
  (\bar\psi\psi\bar\varphi\varphi)^n
  =\exp(\bar\psi\psi+\bar\varphi\varphi)\nn
\end{align}
as in Eq.~\eqref{eq:sumnQ}, where we have used $M=\psi\bar\psi$,
$N=\varphi\bar\varphi$, and $B=\varphi\bar\psi$.  We see that the
transformation \eqref{eq:nf=nc} works and that, unlike in
App.~\ref{app:ex1-div}, we do not have any divergence problem.

\subsection{$N_c=N_b=2$}

In this case we also have $2N_b>N_c$, and therefore divergences would
arise in Eq.~\eqref{eq:cft-alg}.  We now check our result
\eqref{eq:nf=nc} in which no divergence appears.  Using the
parameterization of SU(2) in App.~\ref{app:ex1}, we perform the
integral on the LHS,
\begin{align} 
  \label{eq:e2lhs}
  \int\limits_{{\rm SU}(2)} dU
  \exp\left(\bar{\psi}_{a}^{i}U^{ij}\psi_{a}^{j}+
    \bar{\varphi}_{a}^{i}U^{\dagger ij}\varphi_{a}^{j} \right) =
  \sum_{n=0}^{\infty}\frac{1}{n!(n+1)!}(\det M+\det N+\tr MN)^n
\end{align} 
with $M$ and $N$ given by Eq.~\eqref{eq:mn2} and $B$ and $C$ given by
Eq.~\eqref{eq:bc}.  To do the integral on the RHS, we parameterize
$\mathrm U(2)$ by multiplying the matrix in Eq.~\eqref{eq:SU2}, which
we now call $V$, by a phase $e^{i\phi}$, with $0\leq\phi<2\pi$.  The
corresponding normalized Haar measure is
$dV=(1/4\pi^3)\sin\theta\cos\theta d\theta d\lambda d\eta d\phi$.  We
then obtain
\begin{align} 
  \label{eq:e2rhs}
  \sum_{Q=0}^{\infty}\tilde\chi_Q\int_{\mathrm U(2)} dV & \ 
  {\det}^{-Q}(VB)\exp\left(\bar{\psi}^i_aV_{ab}\varphi^i_b
    +\bar{\varphi}^i_aV_{ab}^\dagger\psi^i_b\right) \nn \\ 
  &=\sum_{n=0}^{\infty}\frac{1}{n!(n+1)!}(\det M+\det N+\tr BC)^n
\end{align} 
with $\tilde\chi_Q$ defined in Eq.~\eqref{eq:chi}.  The two results
agree.  In the derivation of Eqs.~\eqref{eq:e2lhs} and
\eqref{eq:e2rhs} we have used $\det(BC)=\det(MN)$ and
$\tr(BC)=\tr(MN)$.

\subsection{$N_c=2, N_b=1$}
\label{app:e3}

Let us check our result \eqref{eq:nf<nc}.  The LHS is given
by Eq.~\eqref{eq:lhs}.  On the RHS we have
\begin{align}
  &\!\!\!\!\prod_{n=0}^{N_c-N_b-1} \frac{(N_b+n)!}{n!}
  \int_{\mathrm U(N_b)}dV \ {\det}^{N_b-N_c}(VB)
  \exp\left(\bar{\psi}^i_aV_{ab}\varphi^i_b+
    \bar{\varphi}^i_aV_{ab}^\dagger\psi^i_b\right)\nn \\ 
  &=\int_{0}^{2\pi} \frac{\ \rm d \theta\ }{2\pi}\
  (Be^{i\theta})^{-1}\exp\left(Be^{i\theta}+Ce^{-i\theta}\right)\nn\\
  &=\sum_{n=0}^{\infty}\frac{1}{n!(n+1)!}({\bar \psi}^1\psi^1{\bar
    \varphi}^1{\varphi}^1+ {\bar \psi}^2\psi^2{\bar
    \varphi}^2{\varphi}^2+{\bar \psi}^1\psi^2{\bar
    \varphi}^2{\varphi}^1+{\bar \psi}^2\psi^1{\bar
    \varphi}^1{\varphi}^2)^n\:,
\end{align} 
where we have used $B=\varphi^1\bar\psi^1+\varphi^2\bar\psi^2$ and
$C=\psi^1\bar\varphi^1+\psi^2\bar\varphi^2$.  Thus, the RHS agrees
with the result in Eq.~\eqref{eq:lhs}.

\end{document}